\documentclass[%
 reprint,
superscriptaddress,
showkeys,
 aps,
 pra,
]{revtex4-1}

\usepackage[margin=1in]{geometry} 
\usepackage{amsmath,amsthm,amssymb}
\usepackage{graphicx}
\usepackage{subfigure}
\usepackage{braket}
\usepackage{dsfont}
\usepackage{comment}
\usepackage{blkarray}
\usepackage{float}
\makeatletter
\let\newfloat\newfloat@ltx
\makeatother
\usepackage{algorithm}
\usepackage[noend]{algpseudocode}
\usepackage{graphicx}
\usepackage{dcolumn}
\usepackage{bm}
\usepackage{amsfonts}
\usepackage{epsf}  
\usepackage{xcolor}
\usepackage{tikz}
\usepackage{cancel}
\usepackage{multirow}

\usepackage{algorithm}
\usepackage{algpseudocode}
\usepackage{mathtools}

\usepackage{hyperref}
\usepackage[mathlines]{lineno}

\makeatletter

\begin{document}

\preprint{APS/123-QED}

\title{QFold: Quantum Walks and Deep Learning to Solve Protein Folding}

\author{P. A. M. Casares}
 \email{pabloamo@ucm.es}
 \affiliation{Departamento de F\'isica Te\'orica, Universidad Complutense de Madrid.}
\author{Roberto Campos}%
 \email{robecamp@ucm.es}
\affiliation{Departamento de F\'isica Te\'orica, Universidad Complutense de Madrid.}
\affiliation{Quasar Science Resources, SL.}%
\author{M. A. Martin-Delgado}%
 \email{mardel@ucm.es}
\affiliation{Departamento de F\'isica Te\'orica, Universidad Complutense de Madrid.}
\affiliation{CCS-Center for Computational Simulation, Universidad Politécnica de Madrid.}%



\date{\today}

\begin{abstract}

Predicting the 3D structure of proteins is one of the most important problems in current biochemical research. In this article, we explain how to combine recent deep learning advances with the well known technique of quantum walks applied to a Metropolis algorithm. The result, QFold, is a fully scalable hybrid quantum algorithm that, in contrast to previous quantum approaches, does not require a lattice model simplification and instead relies on the much more realistic assumption of parameterization in terms of torsion angles of the amino acids. We compare it with its classical analog for different annealing schedules and find a polynomial quantum advantage, and implement a minimal realization of the quantum Metropolis in IBMQ Casablanca quantum system.

\end{abstract}

\maketitle

\section{\label{sec:intro}Introduction}

Proteins are complex biomolecules, made up of one or several chains of amino acids, and with a large variety of functions in organisms. Amino acids are 20 compounds made of amine $(-NH_2)$ and carboxyl $(-COOH)$ groups, with a side chain that differentiates them. However, the function of the protein is not only determined by the amino acid chain, which is relatively simple to figure out experimentally, but by its spatial folding, which is much more challenging and expensive to obtain in a laboratory. In fact it is so complicated that the gap between proteins whose sequence is known and those for which the folding structure has been additionally analyzed is three orders of magnitude: there are over 200 million sequences available at the UniProt database \cite{uniprot2019uniprot}, but just over 172 thousand whose structure is known, as given in the Protein Data Bank \cite{PDB}. Furthermore, experimental techniques cannot always analyse the tridimensional configuration of the proteins, giving rise to what is called the dark proteome \cite{perdigao2015darkproteome} that represents a significant fraction of the organisms including humans \cite{bhowmick2016darkproteome,perdigao2019dark}. There are even proteins with several stable foldings \cite{bryan2010metamorphicproteins}, and others that have no stable folding called Intrinsically Disordered \cite{dunker2001intrinsically}.

Since proteins are such cornerstone biomolecules, and retrieving their folding so complicated, the problem of protein folding is widely regarded as one of the most important and hard problems in computational biochemistry. 

Until recently, one of the most common approaches to fold proteins was to apply a Metropolis algorithm parameterised in terms of the torsion angles, as is done, for example, in the popular library package Rosetta \cite{Rosetta} and the distributed computing project Rosetta@Home \cite{Rosetta@home,das2007rosetta@home}. The main problem with this approach, though, is that the protein structure prediction problem is combinatorial in nature, and NP-complete even for simple models \cite{hart1997robust,berger1998protein}. For this reason, other approaches are also worth exploring. In the 2018 edition of the Critical Assessment of Techniques for Protein Structure Prediction (CASP) competition \cite{CASP}, for example, the winner was DeepMind's AlphaFold model \cite{AlphaFold}, that was able to show that Deep Learning techniques allow to obtain much better results. DeepMind approach consisted on training a neural network to produce a mean field potential, dependent on the distance between amino acids and the torsion angles, to be later minimized by gradient descent. The latest AlphaFold version further improves the accuracy \cite{AlphaFoldv2}. 

In this article we study how quantum computing could help improve the state of the art in this problem when large error-corrected quantum computers become available. We propose using the prediction of AlphaFold as a starting point for a quantum Metropolis-Hastings algorithm. The Metropolis algorithm is a Markov-chain Monte Carlo algorithm, that is, an algorithm that performs a random walk $\mathcal{W}$ over a given graph. The Metropolis algorithm is specially designed to reach the equilibrium state as quickly as possible, the state $\pi^\beta$ such that $\mathcal{W}\pi^\beta = \pi^\beta$. Slowly  modifying the inverse temperature parameter $\beta$ such that the states with smaller energy become increasingly favoured by the random walk, we should end in the ground state of the system with high probability. Indeed, the intuition why quantum computing might be useful in this problem, is that annealing processes have been widely used to tackle NP-complete problems such as this one \cite{Rosetta, Rosetta@home, das2007rosetta@home}; and its quantization using quantum walks and quantum annealing can decrease the `hitting time' required to find the solution \cite{portugal2013quantum}.


\begin{figure*}[t]
\centering
\includegraphics[width=\textwidth]{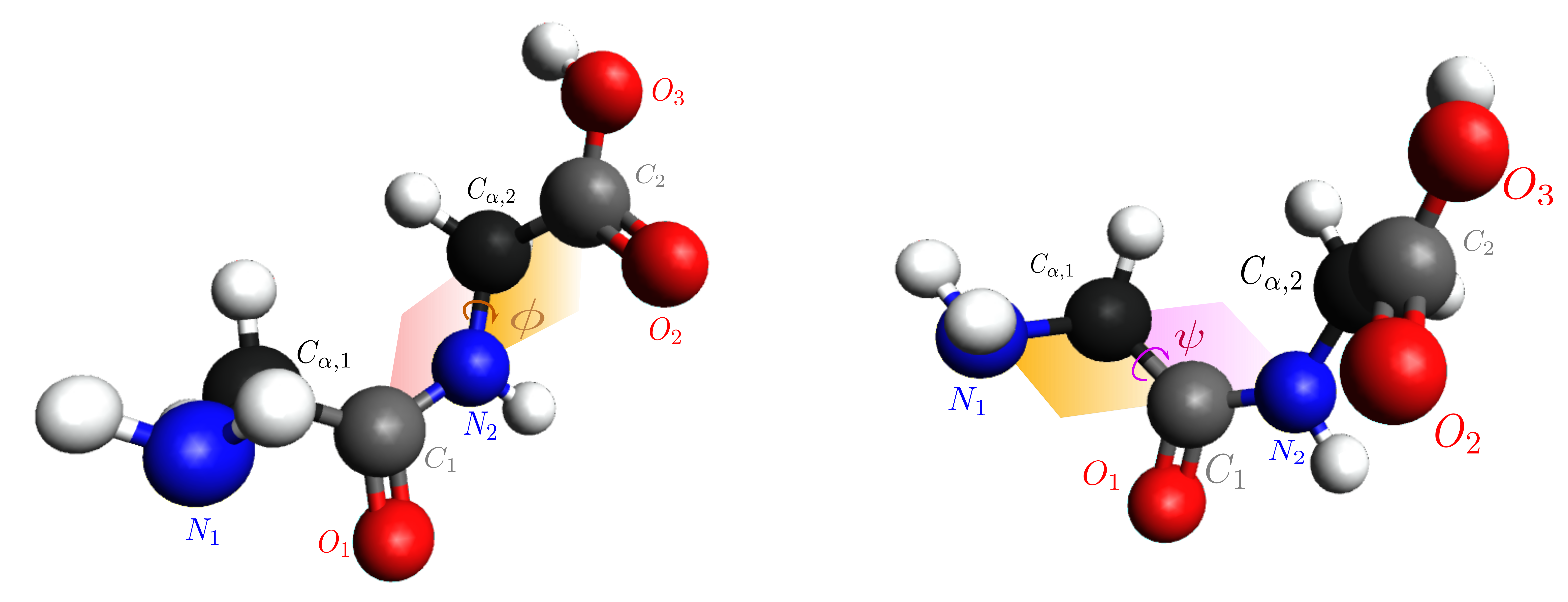}
\caption{Example of the smallest dipeptide: the glycylglycine. Each amino acid has the chain (Nitrogen-$C_{\alpha}$-Carboxy). Different amino acids would have a different side chain attached to the $C_\alpha$ instead of Hydrogen as it is the case for the Glycyne. In each figure we depict either angle $\phi$ or $\psi$. Angle $\psi$ is defined as the torsion angle between two planes: the first one defined by the three atoms in the backbone of the amino acid ($N_1$, $C_{\alpha,1}$, $C_1$), and the second by the same atoms except substituting the Nitrogen in that amino acid by the Nitrogen of the subsequent one: ($C_{\alpha,1}$, $C_1$, $N_2$). For the $\phi$ angle the first plane is made out of the three atoms in the amino acid ($N_2$, $C_{\alpha,2}$, $C_2$) whereas the second plane is defined substituting the Carboxy atom in the amino acid by the Carboxy from the preceding amino acid: ($C_1$, $N_2$, $C_{\alpha,2}$). These graphics were generated using \cite{hanwell2012avogadro} and Inkscape.}
\label{fig:glycylglycine}\end{figure*}


Several modifications of the Metropolis-Hastings algorithm to adapt it to a quantum algorithm have been proposed \cite{wocjan2008speedup,somma2007quantum,somma2008quantum,temme2011quantum,yung2012quantum,lemieux2019efficient}, mostly based on substituting the classical random walk by a Szegedy quantum walk \cite{szegedy2004quantum}. On the contrary, our work takes advantage of the application of a quantum Metropolis algorithm under out-of-equilibrium conditions (e.g. with heuristic annealing schedules) similar to what is usually done classically, and has been done on Ising models \cite{lemieux2019efficient}. Specifically, we aim to simulate this procedure for several small peptides, the smallest proteins with only a few amino acids. We then compare the expected running time with the classical simulated annealing, and also check whether starting from the initial state proposed by an algorithm similar to AlphaFold may speed up the simulated annealing process.

Our work benefits from two different lines of research. The first one makes use of quantum walks to obtain polynomial quantum advantages, inspired mainly by Szegedy work \cite{szegedy2004quantum}, and by theoretical quantum Metropolis algorithms indicated in the above. In contrast with \cite{lemieux2019efficient}, our work focuses only on the unitary heuristic implementation of the Metropolis algorithm, but studies what happens with a different system (peptides) and with different annealing schedules instead of only testing a single linear schedule for the inverse temperature $\beta$. Lastly, we also implement a minimal experimental realization using IBMQ Casablanca (ibmq\_casablanca, processor type Falcon r4) system.

The second line of research related to our work is the use of quantum techniques to speedup or improve the process of protein folding. The literature on this problem \cite{babbush2012construction,robert2021resource, perdomo2012finding, fingerhuth2018quantum, babej2018coarse, perdomo2008construction,outeiral2020investigating, wong2021quantum} and related ones \cite{mulligan2020designing,banchi2020molecular} focuses on such simplified lattice models that are still very hard, and mostly on adiabatic computation. In contrast, our work presents much more realistic fully scalable model, parametrized in terms of the torsion angles. The torsion angles, also called dihedral, are angles between the atoms in the backbone structure of the protein, that determine its folding. An example with the smallest of the dipeptides, the glycylglycine, can be found in figure \ref{fig:glycylglycine}. These angles are usually three per amino acid, $\phi$, $\psi$ and $\omega$, but the latter is almost always fixed at value $\pi$ and for that reason, not commonly taken into account \cite{AlphaFold}. 

What realistic models of protein structure prediction do take into account are the dihedral angles of the side-chains as well as the environment of the protein. While we are limited by the cost of classical simulation of our proof-of-concept quantum algorithm, one of the most remarkable aspects of our proposal is that it allows to incorporate that degree of realism, when provided with a fault-tolerant quantum computer. In such scenario one should substitute the energy evaluation oracle with the quantized (reversible) version of the appropriate Rosetta scoring function, effectively capturing the richness of modern classical methods. This would require expensive and in other situations discouraged arithmetic computations. However, performing such arithmetic computations would amount to a constant multiplicative slowdown that can be offset thanks to the quantum advantage of the quantum Metropolis algorithm. The upside of this procedure is that it allows the algorithm to retain the full precision of classical methods.

\begin{figure*}[t]
\centering
\includegraphics[width=\textwidth]{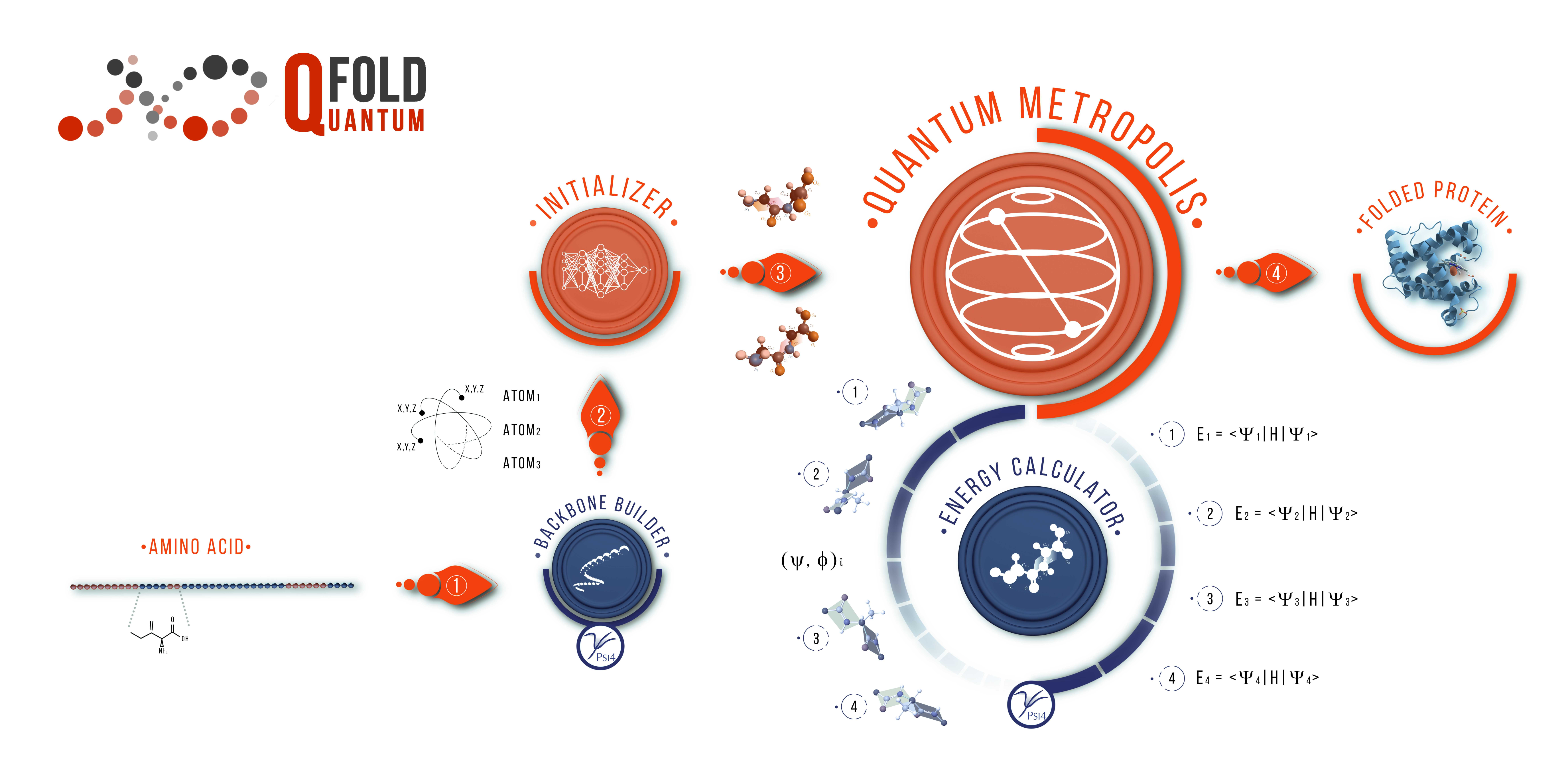}
\caption{Scheme of the QFold algorithm. Starting from the amino acid sequence, we use \textit{Psi4} to extract the atoms conforming the protein, and a \textit{Minifold} module, in substitution of AlphaFold, as initializer. The algorithm then uses the guessed angles by \textit{Minifold} as a starting point (or rather, as the means of the starting von Mises distributions with $\kappa = 1$), and the energy of all possible configurations calculated by \textit{Psi4}, to perform a quantum Metropolis algorithm that finally outputs the torsion angles. In the scheme of the algorithm, the backbone builder represents a subroutine that recovers the links between atoms of the protein, and in particular the backbone chain, using the atom positions obtained from \textit{PubChem} using \textit{Psi4}. The initializer, instantiated in our case by \textit{Minifold}, is a second subroutine that gives a first estimate of the torsion angles, before passing it to the quantum Metropolis. The energy calculator uses \textit{Psi4} to calculate the energy of all possible rotation angles that we want to explore, and these energies are used in the quantum Metropolis algorithm, which outputs the expected folding. For a more detailed flowchart, we refer to the figure \ref{fig:flow_chart}.}
\label{fig:scheme}\end{figure*}

These considerations, and the fact that we use a distilled version of AlphaFold \cite{AlphaFold} as initialization, makes our work different from the usual approach in quantum protein folding: commonly adiabatic approaches have been used so far, whereas our algorithm is digital. The downside of this more precise approach is that the number of amino acids that we are able to simulate is more restricted. 

In summary, the main contributions of our work are threefold: firstly, we design a quantum algorithm that is scalable and realistic, and provided with a fault-tolerant quantum computer could become competitive with current state of the art techniques. Secondly, we analyse the use of different heuristic cooling schedules, and perform ideal quantum simulations of QFold to compare its performance with the equivalent classical Metropolis algorithm, pointing towards a quantum speedup. Thirdly, we carry out a minimal implementation of the quantum Metropolis algorithm in actual quantum hardware.

\section{\label{sec:QFold} QFold algorithm}

\begin{figure*}[t]
\centering
\includegraphics[width=\textwidth]{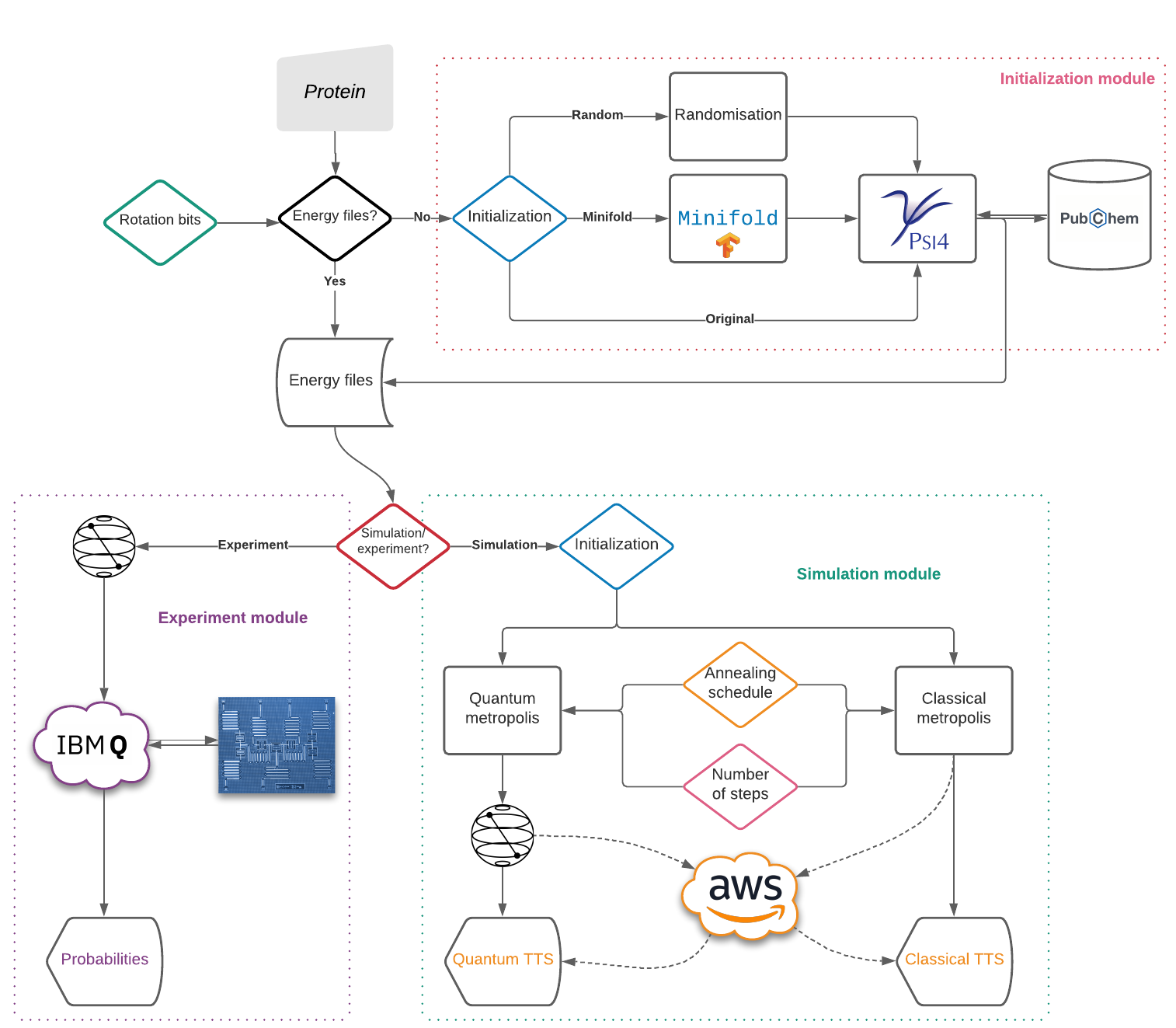}
\caption{Flow chart of the QFold algorithm. This figure has to be viewed with the help of figure \ref{fig:scheme}. QFold has several functionalities integrated altogether, that could be summarized in an initialization module, a simulation module, and an experiment module. We denote by diamonds each of the decisions one has to make. The top part constitutes the initialization module, where \textit{Minifold} can be used to get a guess of the correct folding, and \textit{Psi4} and the original geometry from \textit{PubChem} are used to calculate the energies of rotations. The bottom half represents the \texttt{experiment} or \texttt{simulation} algorithms, that output either probabilities or quantum/classical TTS of the corresponding metropolis algorithms, making use of \textit{Qiskit}. 
}
\label{fig:flow_chart}\end{figure*}

The algorithm we introduce is called QFold and has three main components that we will introduce in this section (see figure \ref{fig:scheme} for a scheme of QFold): an initialization routine to find a good initial guess of the dihedral angles that characterise the protein folding, a quantum Metropolis to find an even lower energy state from the initial guess, and a classical metropolis to compare against the quantum Metropolis to assess possible speedups. The aim of this section is to introduce the theoretical background we have used for our results.

\subsection{\label{sec:Initialization} Initializer}

QFold makes use of quantum walks as a resource to accelerate the exploration of protein configurations (see figures \ref{fig:scheme} and \ref{fig:flow_chart}).
However, in nature proteins do not explore the whole exponentially large space of possible configurations in order to fold. In a similar fashion, QFold does not aim to explore all possible configurations, but rather uses a good initialization guess state based on Deep Learning techniques such as AlphaFold. Since such initial point is in principle closer to the actual solution in the real space, we expect it to be most helpful the larger the protein being modelled. In fact, one of the motivations for our work was the fact that adding a Rosetta relaxation at the end of the AlphaFold algorithm was able to slightly improve the results of the AlphaFold algorithm \cite{AlphaFold}. Notice that a Rosetta relaxation is the way Rosetta calls its classical Metropolis algorithm. Therefore, we expect that improved versions of Rosetta, using in our case quantum walks, could be of help to find an even better solution to the protein structure prediction problem than the one provided by only using AlphaFold.

The AlphaFold initializer starts from the amino acid sequence ($S$), and performs the following procedures:
\begin{enumerate}
    \item First perform a Multiple Sequence Alignment (MSA) procedure to extract features of the protein already observed in other proteins whose folding is known. 
    \item Then, parametrizing the proteins in terms of their backbone torsion angles $(\phi, \psi)$ (see figure \ref{fig:glycylglycine}), train a residual convolutional neural network to predict distances between amino acids, or as they call them, residues.
    \item Train also a separate model that gives a probability distribution for the torsion angle conditional on the protein sequence and its previously analysed MSA features, $P(\phi,\psi| S, MSA(S))$. This is done using a 1-dimensional pooling layer that takes the predicted distances between amino acids and outputs different secondary structure such as the $\alpha$-helix or the $\beta$-sheet \footnote{The $\alpha$-helix and the $\beta$-sheet correspond to two common structures found in protein folding. Such structures constitute what is called the secondary structure of the protein, and are characterised because $(\phi, \psi)= (-\pi/3, -\pi/4)$ in the $\alpha$-helix, and $(\phi, \psi)= (-3\pi/4, -3\pi/4)$ in the $\beta$-sheet, due to the hydrogen bonds that happen between backbone amino groups NH and backbone carboxy groups CO}. To make the prediction, the algorithm makes use of bins of size 10º, effectively discretising its prediction.
    \item All this information, plus some additional factors extracted from Rosetta, is used to train an effective potential that aims to give smaller energies to the configurations that the model believes to be more likely to happen in nature.
\end{enumerate}

Finally, at inference time one starts from a rough guess using the MSA sequence and performs gradient descent on the effective potential. One can also perform several attempts with noisy restarts and return the best option. Interestingly enough, the neural network is also able to return an estimation of its uncertainty. Such uncertainty is measured by the parameter $\kappa$ in the von Mises distribution and plays the role of the inverse of the variance. The von Mises distribution is the circular analog of the normal distribution, and its use is justified because angles are periodic variables \cite{von2014mathematical}. 


\subsection{\label{sec: Classical Metropolis} Classical Metropolis}

As we have mentioned in the introduction, a relatively common approach to perform protein structure prediction has been to use the Metropolis algorithm. The Metropolis algorithm is an algorithm that performs a random walk over the configuration space $\Omega$.
The configuration space is the abstract space of possible values the torsion angles of a given protein can take. As such, a given state $i$ is a list of values for such torsion angles. In particular, for computational purposes, we will set that angles can take values from a given set, that is, it will not be a continuous but a discrete distribution. Over such space we can define probability distributions. Furthermore, since those angles will dictate the position of the atoms in the protein, the state $i$ will also imply an energy level $E_i$, due to the interaction of the atoms. In the Rosetta library, the function that calculates an approximation to such energy is called the scoring function.

Starting from a state $i$, the Metropolis algorithm proposes a change uniformly at random to one of the configurations, $j$, connected to $i$. We will call $T_{ij}$ to the probability of such proposal. Then this change is accepted with probability
\begin{equation}
    A_{ij}= \min \left(1,e^{-\beta(E_j-E_i)}\right),
    \label{Metropolis update probability}
\end{equation}
resulting in an overall probability of change $i \rightarrow j$ at a given step $\mathcal{W}_{ij} = T_{ij} A_{ij}$. One often works with a ensemble of points $\{\mathbf{x}^t\}$ that get individually updated to $\{\mathbf{x}^{t+1}\}$, a sample from the probability distribution defined by the metropolis algorithm at $\beta$.

Slowly varying it one decreases the probability that steps that increase the energy of the state are accepted, and as a consequence when $\beta$ is sufficiently large, the end state is a local minima. If this annealing procedure is done sufficiently slowly, one can also ensure that the minima found is the global minima. However, in practice one does not  perform this annealing as slowly as required, resorting instead to heuristic restarts of the classical walk, and selecting the best result found by several different trajectories.

In our implementation we emulate having oracle access to the energies of different configurations. We give details of our oracle implementation in section \ref{sec:Experiment&simulation_conditions}.

\subsection{\label{sec: Quantum Metropolis} Quantum Metropolis}

A natural generalisation of the Metropolis algorithm explained in the previous section is the use of quantum walks instead of random walks. The best known quantum walk for this purpose is Szegedy's \cite{szegedy2004quantum}, that consists of two rotations similar to the rotations performed in Grover's algorithm \cite{Grover}. Szegedy's quantum walk is defined on a bipartite graph. Given the acceptance probabilities $\mathcal{W}_{ij} = T_{ij}A_{ij}$, $A_{ij}$ defined in \eqref{Metropolis update probability}, for the transition from state $i$ to state $j$, one defines the unitary
\begin{equation}
    U\ket{j}\ket{0} := \ket{j} \sum_{i \in \Omega} \sqrt{\mathcal{W}_{ji}}\ket{i} = \ket{j}\ket{p_j}.
\end{equation}
Taking 
\begin{equation}
    R_0 := \mathbf{1} - 2\Pi_{0}= \mathbf{1} - 2(\mathbf{1}\otimes \ket{0}\bra{0})
\end{equation}
the reflection over the state $\ket{0}$ in the second subspace, and $S$ the swap gate that swaps both subspaces, we define the quantum walk step as 
\begin{equation}
 W:= U^\dagger S U R_0 U^\dagger S U R_0. \label{W definition}
\end{equation}
We refer to Appendix \ref{sec:Szegedy} and figure \ref{fig:quantum_walk} for a detailed account on the use of these quantum walks in a similar way to the Grover rotations. For completeness, in Appendix \ref{sec:Szegedy} we review the theoretical basis of Szegedy quantum walks that
leads us to believe that a quantum advantage is
possible in our problem.

It is well known that if $\delta$ is the eigenvalue gap of the classical walk, and $\Delta$ the phase gap of the quantum walk, then the complexity of the classical walk is $O(\delta^{-1})$, the complexity of the quantum algorithm $O(\Delta^{-1})$, and the relation between the phase and eigenvalue gap is given by $\Delta = \Omega(\delta^{1/2})$ \cite{magniez2011search}. Our algorithm aims to explore what is the corresponding advantage in practice. 

The quantum Metropolis algorithm that we employ \cite{lemieux2019efficient} is based on a modification of the Szegedy quantum walk, substituting the bipartite graph by a coin. The change from one quantum walk to another is a conjugation by an isomorphism that does not change the theoretical results (see appendix \ref{sec:OOE-Metropolis_algorithm}). We will have 3 quantum registers: $\ket{\cdot}_S$ indicating the current state of the system, $\ket{\cdot}_M$ that indexes the possible moves one may take, and $\ket{\cdot}_C$ the coin register. We may also have ancilla registers $\ket{\cdot}_A$. The quantum walk step is defined as half the Szegedy quantum walk \cite{lemieux2019efficient}
\begin{equation}
    \tilde{W} = R V^\dagger B^\dagger F B V.
    \label{quantum walk operator tilde W}
\end{equation}
Here, $V$ prepares a superposition over all possible steps one may take in register $\ket{\cdot}_M$, $B$ rotates the coin qubit $\ket{\cdot}_C$ to have amplitude of $\ket{1}_C$ corresponding to the acceptance probability indicated by \eqref{Metropolis update probability}, $F$ changes the $\ket{\cdot}_S$ register to the new configuration (conditioned on the value of $\ket{\cdot}_M$ and $\ket{\cdot}_C = \ket{1}_C$), and $R$ is a reflection over the state $\ket{0}_{MCA}$. 

Although other clever options are available \cite{lemieux2019efficient}, here we implement the simplest heuristic algorithm, which consists of implementing $L$ of such steps
\begin{equation}
    \ket{\psi(L)} := \tilde{W}_L ... \tilde{W}_1 \ket{\pi_0},
    \label{L quantum walk steps}
\end{equation}
where $t = {1,...,L}$ also defines an annealing schedule, for chosen values of $\beta(t)$ at each step. More detailed explanation of the algorithm \cite{lemieux2019efficient} can be found in appendix \ref{sec:OOE-Metropolis_algorithm}.

\section{\label{sec:Metrics} Figures of merit}

When looking for a metric to assess the goodness of given solutions to protein structure prediction we have to strike a balance between two aspects: on the one hand, we want a model that with high probability finds the correct solution. On the other hand, we would like such procedure to be fast. For example going through all configuration solutions would be quite accurate, albeit extremely costly.

A natural metric to use in this context is then the Total Time to Solution (TTS) \cite{lemieux2019efficient} defined as the average expected time it would take the procedure to find the solution if we can repeat the procedure in case of failure:
\begin{equation}
    TTS(t):= t \frac{\log (1-\delta)}{\log ( 1-p(t))}.
    \label{TTS}
\end{equation}
where $t\in \mathbb{N}$ is the number of quantum/random steps performed in an attempt of the quantum/classical Metropolis algorithm, $p(t)$ the probability of hitting the right state after those steps in each attempt, and $\delta$ a target success probability of the algorithm taking into account restarts, that we set to the usual value of $0.9$. Since the latter is a constant, the value of $TTS(t)$ with other value of $\delta$ is straightforward to recover. This metric represents the compromise between longer walks and the corresponding expected increase in the probability of success. In this figure, the way we have to compare classical and quantum walks is to compare the minimum values achieved, the $\min_t TTS(t)$. Similar metrics have also been defined previously \cite{albash2018demonstration}.

On the other hand, there is a small modification of the classical algorithm that could improve its TTS, because we only output the last configuration of the random walk instead of the state with the minimum energy found so far, a common choice for example in the Rosetta@Home project. The reason for not having included this modification is because the length of the classical path, 2 to 50 steps, represents a sizable portion of the total space that ranges from 64 to 4096 available configurations, what will not be the case for large proteins. We believe that had we run the classical experiments with that modification, we would have introduced a relatively large bias in the results, favouring the classical random walks in the smallest instances of the problem, and therefore likely overestimating the quantum advantage.

For the experiment run in IBM Quantum systems and whose results can be found in section \ref{sec:Hardware_experiment_results}, the metric we use instead of the TTS is the probability of measuring the correct answer, and in particular whether we are able to detect small changes in the probability corresponding to the correct solution. Measuring the TTS here would not be interesting due to the high level of noise of the circuit.

\section{\label{sec:Experiment&simulation_conditions}Simulation setup}

For the assessment of our algorithm we have built a simulation pipeline that allows to perform a variety of options. The main software libraries used are \textit{Psi4} for the calculation of energies of different configurations in peptides \cite{turney2012psi4}, a distilled unofficial version of AlphaFold dubbed \textit{Minifold} \cite{ericalcaide2019minifold}, and \textit{Qiskit} \cite{Qiskit} for the implementation of quantum walks.

The most expensive part of the algorithm is the `oracle' that computes the acceptance probabilities. Instead of compiling the Rosetta scoring function to a quantum circuit \cite{Rosetta}, that we leave for future work, we have classically precomputed all the energies and encoded a true oracle. Obviously, this is not feasible for larger cases but allows to model the behaviour of the algorithm. The precalculation was carried out with the common 6-31G basis functions \cite{jensen2013atomic}, and the procedure of Moller-Plesset to second order \cite{helgaker2014molecular} as a more accurate and not too expensive alternative to the Hartree-Fock procedure.

On the other hand, we also need to select the number of bits that specify the discretization of the torsion angles. For example, 1 bit means that angles can take values in $\{0, \pi\}$, whereas 2 bits indicate discretization in $\pi/2$ radians. Notice that the precision of 10º of AlphaFold when reporting their angles, that we indicated in section \ref{sec:Initialization}, is intermediate between $b = 5$ and $b = 6$. Increasing the discretization size makes the simulation more precise but also more expensive, as we will see in figure \ref{fig:fixed_beta_TTS_slope}.

One advantage of our proposal is that the uncertainty of the deep learning module prediction might be included in the initialization. In fact, if the original AlphaFold algorithm were to be used, the $\kappa$ values returned by AlphaFold could actually be used instead of our default value $\kappa = 1$. The amplitudes corresponding to the von Mises probability distribution could be efficiently encoded using the Grover-Rudolph state preparation procedure \cite{State_prep_grover}.

We have implemented and tested several options for the annealing schedule of $\beta$. The implemented schedules are:
\begin{itemize}
\begin{subequations} \label{Annealing schedules}
    \item \texttt{Boltzmann} or \texttt{logarithmic} implements the famous logarithmic schedule \cite{kirkpatrick1983optimization}
    \begin{equation}
        \beta(t) = \beta(1) \log(t e) = \beta(1) \log(t) + \beta(1).
    \end{equation}
    Notice that the multiplication of $t$ times $e$ is necessary in order to make a fair comparison with the rest of the schedules, so that they all start in $\beta(1)$.
    \item \texttt{Cauchy} or \texttt{linear} implements a schedule given by
    \begin{equation}
        \beta(t) = \beta(1) t.
    \end{equation}
    \item \texttt{geometric} defines
    \begin{equation}
        \beta(t) = \beta(1) \alpha^{-t+1},
    \end{equation}
    where $\alpha<1$ is a parameter heuristically set to $0.9$. 
    \item And finally \texttt{exponential} uses
    \begin{equation}
        \beta(t) = \beta(1) \exp(\alpha (t-1)^{1/N}),
    \end{equation}
    where $\alpha$ is again set to $0.9$ and $N$ is the space dimension, which in this case is equal to the number of torsion angles. 
\end{subequations}
\end{itemize}
For comparison purposes, the value of $\beta(1)$ chosen has been heuristically optimized to $50$.

\begin{table*}[]
\begin{tabular}{|c|c|c|c|c|c|}
\hline
    Peptides              &    Precision random               &      Precision minifold             & $b$  & \texttt{quantum min(TTS) random} &  \texttt{quantum min(TTS) minifold}  \\ \hline \hline
\multirow{3}{*}{Dipeptides} & \multirow{3}{*}{ 0.53} & \multirow{3}{*}{ 0.53} & 3 &  \textbf{136.25}&  270.75\\ \cline{4-6}
                  &                   &                   & 4   & \textbf{547.95} &   1137.45\\ \cline{4-6}
                  &                   &                   & 5  & \textbf{1426.28} &  1458.02 \\ \hline
            Tripeptides      &   0.46                &     \textbf{0.71}          &  2  & 499.93  &  \textbf{394.49} \\\hline
             Tetrapeptides     &    0.51               &     \textbf{0.79}             & 1  & 149.80  & \textbf{26.30}\\ \hline
\end{tabular}
\caption{Table of average precisions defined in equation \eqref{precision definition}, and corresponding quantum minimum TTS, defined in equation \eqref{TTS} as the expected number of steps it would take to find the solution using the quantum algorithm, with different initializations. $b$ denotes the rotation bits, and in bold we have indicated which of minifold or random values are best. The aim of this table is understanding the impact of minifold initialization in the quantum $\min TTS$, our figure of merit. The two main aspects to notice from the table are that \textit{Minifold} precision grows with the size of the peptide, and that when it is the case that the minifold precision is higher, the corresponding quantum min TTS values are lower than their random counterparts. This supports the idea that using a smart initial state helps to find the native folding of the protein faster.}
\label{table:precision}
\end{table*}

The amount of Random Access Memory of the simulator is the main constraint we face to scale up the size of a simulation. The reason is that our more straightforward but rather inefficient way of computing the acceptance probabilities has a bad scaling: it has to iterate over all possible values of the angles and proposed movements. However, the alternative would require significant circuit compilation, and for that reason it is left for future work.
Due to this constraint we have simulated, for a fixed value of $\beta$:
\begin{itemize}
    \item Dipeptides with 3 to 5 rotation bits.
    \item Tripeptides with 2 rotation bits.
    \item Tetrapeptides with a single rotation bit.
\end{itemize}
Additionally, dipeptides with 6 bits, tripeptides with 3 bits and tetrapeptides with 2 bits can be simulated for a few steps, but not enough of them to calculate our figure of merit with confidence.
If the $\beta$ value is not fixed but follows some annealing schedule, the requirements are larger, but we can still simulate the same peptides.
Further than that, the Random Access Memory requirements for an ideal (classical) simulation of the quantum algorithm become unmanageable with our resources. Notice that Qiskit simulator supports 32 qubits at the moment, but our system is more constrained by the depth of the circuit, which can run into millions of gates. 

\section{\label{sec:Results} Simulation results}

\subsection{\label{sec:Initialization results}Initialization}

In this section we analyse the impact of different initialization methods for the posterior use of quantum/classical walks.
As an initializer, we decided to use (and minorly contributed to) \textit{Minifold} because even though it does not achieve the state of the art in the prediction of the angles, it is quite simple and sufficient to illustrate our point. \textit{Minifold} \cite{ericalcaide2019minifold} uses a residual network implementation given in Tensorflow and Keras \cite{abadi2016tensorflow, chollet2015keras}. Perhaps the most important detail of using this model is that because we are trying to predict small peptides, and \textit{Minifold} uses a window of 34 amino acids for its predictions, we had to use padding. Another detail is that while \textit{Minifold} did not use MSA techniques because we were dealing with small peptides, when working with larger proteins the model should necessarily use them to obtain good precision estimates.

The metrics that we analyse in this case are twofold: in the first place, we would like to see whether \textit{Minifold} achieves a better precision on the angles than random guessing. This is a necessary condition for our use of \textit{Minifold}, or more generally, any smart initialization module, to make sense. We measure the precision as:
\begin{equation}
    p = 1-\frac{d(\alpha, \tilde{\alpha})}{\pi},
\label{precision definition}
\end{equation}
where $\tilde{\alpha}$ is the estimated angle (either $\phi$ or $\psi$) given by \textit{Minifold} or chosen at random, $\alpha$ is the true value calculated from the output of \textit{Psi4} and \textit{PubChem} (see figure \ref{fig:glycylglycine}), and $d$ denotes the angular distance. 

In table \ref{table:precision} there is a summary comparing the average precision results of minifold and random initialization broken down by the protein and bits. The dipeptide results show that due to the small size of the peptide, the former initialization has barely better precision. However, this situation improves for tripeptides and tetrapeptides, getting a better precision, and as a consequence, lower TTS values.

\subsection{\label{sec:Fixed_beta_results}Fixed $\beta$}

\begin{figure}[t]
\includegraphics[width=.97\textwidth/2]{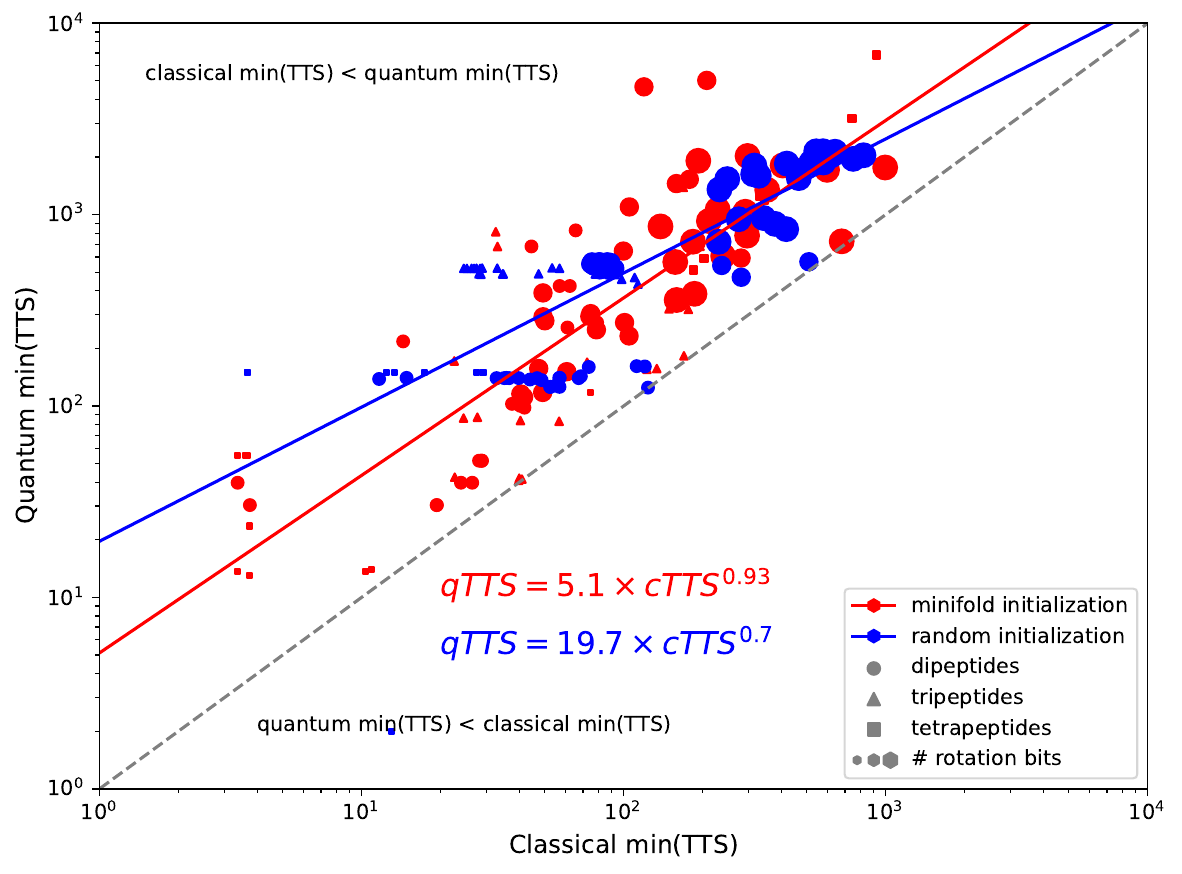}
\centering
\caption{Comparison of the classical and quantum minimum TTS achieved for the simulation of the quantum Metropolis algorithm with $\beta = 10^3$, for 10 dipeptides (with $b = 3, 4, 5$ rotation bits of precision in the angles), 10 tripeptides ($b = 2$) and 4 tetrapeptides ($b = 1$), also showing the different initialization options (random or minifold), and the best fit lines. In dashed grey line we separate the space where the quantum TTS is smaller than the classical TTS. 
The key aspect to notice in this graph is that although for smaller instances the quantum algorithm does not seem to match or beat the times achieved by the classical Metropolis, due to the exponent being smaller than one (either $0.89$  or $0.53$ for minifold or random respectively) for average size proteins we can expect the quantum advantage to be dominant and make the quantum Metropolis more useful than its classical counterpart. In sections \ref{sec:Initialization results} and \ref{sec:Fixed_beta_results} we explain why the random initialization exponent seems more favourable than the minifold exponent and discuss further details respectively.
}
\label{fig:fixed_beta_TTS_slope}\end{figure}

\begin{figure*}[t]
\centering
\makebox[\textwidth][c]{\includegraphics[width=1.\textwidth]{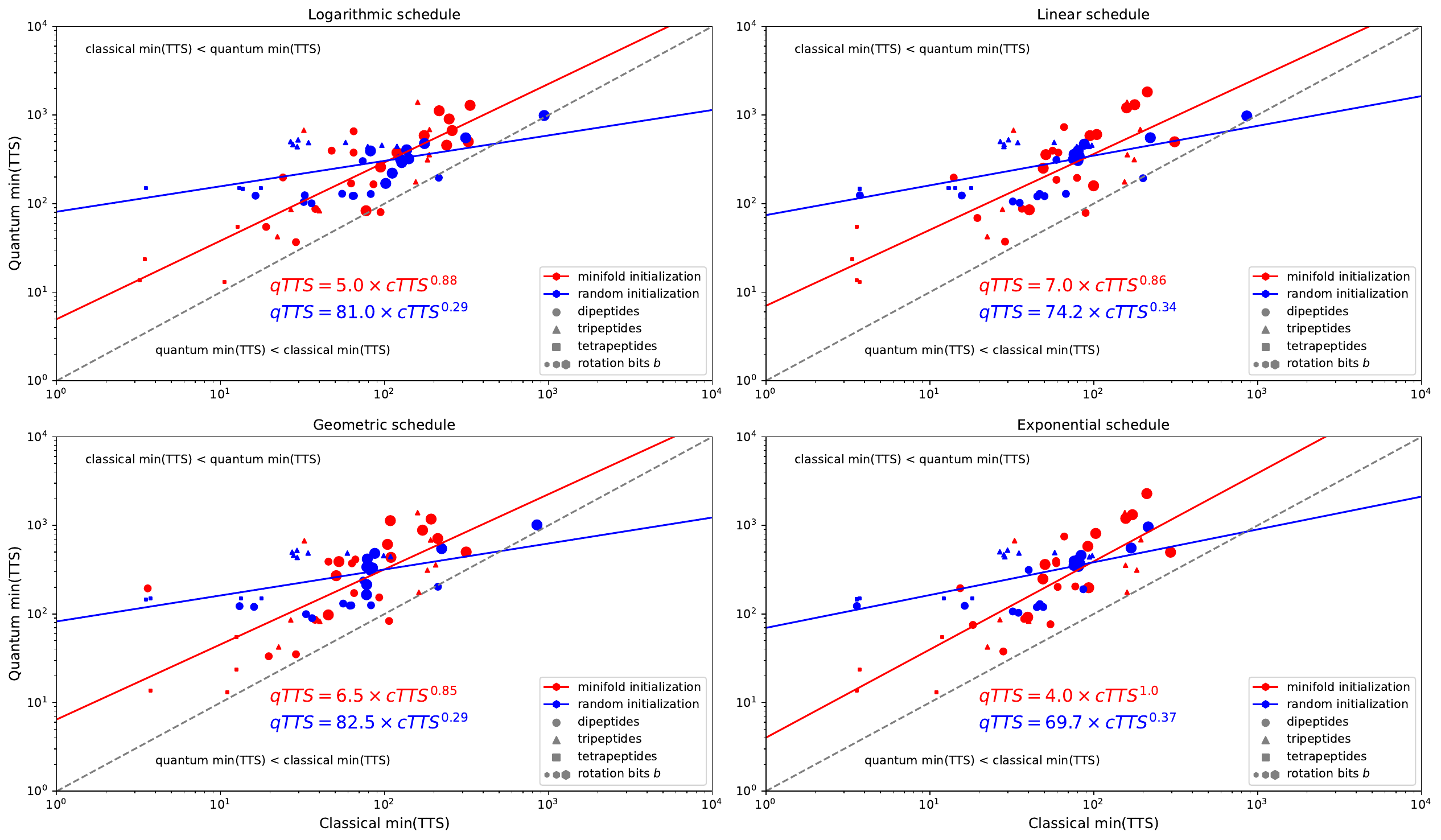}}
\caption{Comparison of the classical and quantum minimum TTS achieved for the same peptides as those in figure \ref{fig:fixed_beta_TTS_slope}, except that due to computational cost we do not include dipeptides with 5 rotation bits. This figure corresponds to section \ref{sec:Variable_beta_results}, and shows the different initialization options (random or minifold) and annealing schedules (\texttt{Boltzmann/logarithmic}, \texttt{Cauchy/linear}, \texttt{geometric} and \texttt{exponential}), and the best fit lines. In dashed grey line we depict the diagonal. The corresponding fit exponents are given in table \ref{table:exponents}, where in three out of the four cases using an annealing schedule increases the quantum advantage. On the other hand, using an exponential schedule does not seem to give but a tiny advantage when used with a minifold initialization.}
\label{fig:var_beta_TTS_slope}\end{figure*}

If \textit{Minifold} having a greater precision in the angles than random guessing was a precondition for our analysis to make sense, the actual metric we are interested in is whether it has some impact reducing the TTS metric. Otherwise we could avoid using an initialization module altogether. Since the expected quantum advantage is polynomial in nature we are interested in finding the exponent that expresses it. There will be a quantum advantage when the quantum minimum TTS grows as a power of the classical minimum TTS with exponent lower than 1. For example, a quadratic quantum advantage implies such exponent being 0.5. To compute this exponent we use the standard technique of linear least square fitting in the logarithmic scale for both $x$ and $y$ variables, the classical and quantum minimum TTS achieved respectively. This fit is shown in figures \ref{fig:fixed_beta_TTS_slope} and \ref{fig:var_beta_TTS_slope}.

In the first of those figures, the exponent of random initialization model is smaller than the one corresponding to the minifold initialization in the fit. While this may seem to indicate that our initialization is harmful to the convergence of the quantum algorithm, in fact the explanation is quite the opposite: for the smaller instances of the problem, and very specially in the case of random initialization, the minimum TTS value is achieved for $t=2$ as can be seen from the two horizontal structures formed by the blue points in the figure, meaning that in such cases only using the minifold initialization the quantum algorithm is able to profit from the incipient quantum advantage. The reason is that in the smallest instances, the constant multiplicative prefactors inherent to the quantum walk have larger effect than the eigenvalue gap that governs the quantum advantage. As a consequence, the system is faster just repeatedly choosing randomly than carrying out the quantum walk.

This effect disappears for larger instances of the problem, but while for random initialization there is a penalisation of the quantum minimum TTS in the smallest problems (thus lowering the exponent), the minifold initialization is capable of correcting for this effect, lowering the TTS of smaller instances of the problem and, as a bonus, rising the exponent. We are therefore inclined to believe that the minifold exponent more accurately represents the true asymptotic exponent of the algorithm for this problem. In conclusion, while the small size of our experiments does not allow us to see the benefits of using a smart initialization, they have been important to get a calibrated estimate of the actual quantum advantage, and we can also see that it helps reduce the TTS cost both in the classical and quantum algorithm, which nicely fits the intuition that being closer-than-random to the solution helps find the solution faster.

We now discuss whether we are able to observe a quantum advantage in the TTS, our figure of merit. We will again be discussing the results given in figure \ref{fig:fixed_beta_TTS_slope}, for it represents the best fit to the classical vs quantum TTS, and therefore accurately depicts the expected quantum advantage: the slopes separated for the different initialization options are $0.89$ for the minifold initialization and $0.53$ for the random initialization. As a consequence, if these trends are sustained with larger proteins, there is a polynomial advantage. 

The final question is what does this advantage mean for the modelling of large enough proteins. For that, we only need two additional ingredients: \textit{how the expected classical $\min TTS$ scales with the size of the configuration of the problem} and the error correction overhead. On the first question, our data in this respect is even more restricted because we only have access to configuration spaces of 64, 256 and 1024 angle configurations. Therefore we are making a regression to only three points, but could give us nevertheless some hints of whether our technique, the use of quantum Metropolis algorithm, will be helpful to solve the tridimensional structure of proteins given a large enough and fault tolerant quantum computer. The regression exponent of a $\log(size)$ vs $\log(\text{classical } \min TTS)$ fit using both random and minifold initializations is $r = 0.88$, which should not be confused with those in figure \ref{fig:fixed_beta_TTS_slope}. 

Let us take, for the sake of giving an example, an average 250 amino acids protein, which has approximately 500 angles to fix. If we use $b = 6$ bits to specify such angles as might be done in a realistic setting, the classical $\min_t TTS$ would be $\approx(2^b)^{2\times 250\times r} = (2^6)^{500\times 0.88}$. The quantum $\min_t TTS$, on the other hand, will be such number to the corresponding exponent of figure \ref{fig:fixed_beta_TTS_slope},  that we will call $e_m = 0.89$ and $e_r = 0.53$ for minifold and random. This will translate to a speedup factor of between $\approx 10^{87}$ and $10^{373}$, although the latter represents an upper bound and the former is probably closer to the actual value. 

However, in the previous estimates we did not take into account the quantum to classical overhead due to error correction. While doing so accurately would require to compile a quantum circuit for the Rosetta scoring function, we can nevertheless give some back-of-the-envelope estimates based on the literature. For example, it is known that performing a NAND logic gate would take approximately 10 orders of magnitude difference using early fault-tolerant quantum computers and classical implementations \cite{babbush2021focus}. Similarly, for the problem of performing simulated annealing on a small Sherrington-Kirkpatrick model, the time gap for performing a single update step would be at least $t_Q/t_C \geq 6.3\cdot 10^7$ \cite{babbush2021focus}. Our problem is more complex and therefore these estimates should be treated as lower bounds.

To make a more precise account of this overhead, let us estimate the cost of performing a 64-bit multiplication in both a quantum and a classical computer. Notice that since the Rosetta scoring function is made out of a constant number of similar operations, the multiplicative slowdown will approximately be the slowdown in one of those operations. Let us start with the classical case. Multiplying two $n$-bit numbers can be done with $n^2$ bit additions, and a single full-adder will be composed of 2 XOR, 2 AND and 1 OR gates. Assuming that each of these take similar time as the NAND gate \cite{babbush2021focus}, the cost of a full adder would be $5\cdot 10^{-9}$ transistor seconds. The total cost of a multiplication of two 64-bit numbers is then $\approx 2\cdot 10^{-5}$ transistor seconds.

The cost of a $n$-bit multiplication in a quantum computer is $21n^2$ T-gates \cite{munoz2017t} (integer division has a similar cost of $14n^2$ \cite{thapliyal2017quantum}). Assuming a same order of magnitude synthesis time for a Toffoli and a T-gate (10 qubit seconds \cite{babbush2021focus}), in this case, the time to perform the multiplication would be $\approx 10^6$ qubit seconds. Consequently, we can see $\approx 11$ orders of magnitude difference between the classical and quantum time required to perform a step.

This reveals that even if we take into account the slowdown due to error correction and other factors in quantum computing, the use of the quantum Metropolis would be competitive. And this conclusion is robust: if we took the quantum advantage exponent to be just $e = 0.95$ and the growth of the TTS with the size of the space an extremely conservative $r =0.5$, the quantum speedup before factoring in the operating frequency of the computer will still be a factor of $\approx 10^{22}$. Larger proteins, which exist in nature, will exhibit even larger speedups in the TTS.


The previous estimates reaffirm the fact that to profit from the quantum advantage we found in our simulations in Fig \ref{fig:fixed_beta_TTS_slope}, we need the use of a fully-fledged quantum computer with achieved fault tolerance. They also highlight the convenience to carry out further research in different variants of the quantum metropolis algorithm achieving higher speedups \cite{babbush2021focus}.

\subsection{\label{sec:Variable_beta_results} Annealing schedules and variable $\beta$}

In the previous section we analysed what quantum advantage could be expected when using a fixed $\beta$ schedule, resulting in a \textit{standard quantum walk} with several steps. However, Metropolis algorithms are rarely used in practice with a fixed $\beta$, since at the beginning of the algorithm one would like to favour exploration, requiring a low $\beta$ value, and at the end one would like to focus mostly on exploitation, that is achieved with a high $\beta$ value. This necessity is linked  to the well known exploration-exploitation trade-off in Machine Learning and more generally in Optimization \cite{sutton}.

\begin{table}[t]
\centering
\begin{tabular}{ |c||c|c|  }
 \hline
 \multicolumn{3}{|c|}{Fit exponents} \\
 \hline
 Schedule & Random initial. & Minifold initial.\\
 \hline
 Fixed $\beta$   &$0.70\pm 0.08$  & $0.93\pm 0.06$\\
 Logarithmic &   $0.29\pm 0.07$  & $0.88\pm 0.09$\\ 
 Linear &$0.34\pm0.07$  & $0.86\pm 0.11$\\
 Exponential    &$0.37\pm 0.07$  & $1.00\pm 0.12$\\
 Geometric &   $0.29\pm 0.07$  & $0.85\pm 0.18$\\
 \hline
\end{tabular}\caption{Table of scaling exponents for different annealing schedules and initialization options. The peptides are the same, except that for fixed $\beta$ we have also included dipeptides with 5 bits of precision, what is costly for the rest of schedules. For fixed $\beta$, the value heuristically chosen was $\beta = 1000$, while the initial $\beta$ value in each of the schedules, defined in \eqref{Annealing schedules}, is $\beta(1) = 50$. The uncertainty is expressed via the standard deviation in the expected exponent, calculated with the bootstrapping method \cite{efron1992bootstrap}.}
\label{table:exponents}
\end{table}

The annealing schedule with the strongest theoretical background is usually called the inhomogeneous algorithm, and it is known because one can prove that the algorithm will converge to the global minimum value of the energy with probability 1, albeit generally too slowly (section 3.2 \cite{van1987simulated}). Its implementation is conceptually similar to the Boltzmann or logarithmic schedule that we use, but with a different prefactor \cite{van1987simulated}.

Therefore, the question we want to answer in this section is what happens when we use our quantum Metropolis algorithm outside of the equilibrium, that is, when we use a schedule that changes faster than the theoretical inhomogeneous algorithm. For this task, several optimization schedules have been proposed, of which we have implemented and tested four different options whose mathematical formulation can be seen in \eqref{Annealing schedules}.

Our conclusions from figure \ref{fig:var_beta_TTS_slope} and table \ref{table:exponents} are that using a variable schedule, the quantum advantage can be made larger than that of the fixed temperature algorithm. However, not all cases give the same advantage, the \texttt{exponential} schedule giving practically none, and the \texttt{geometric} schedule explained in section 5.2 of \cite{van1987simulated} being the most promising, with an exponent of $0.85$ if minifold initialization is used. \texttt{linear} and \texttt{logarithmic} schedules lie in between.

\begin{figure*}[t]
\centering
\includegraphics[width=\textwidth]{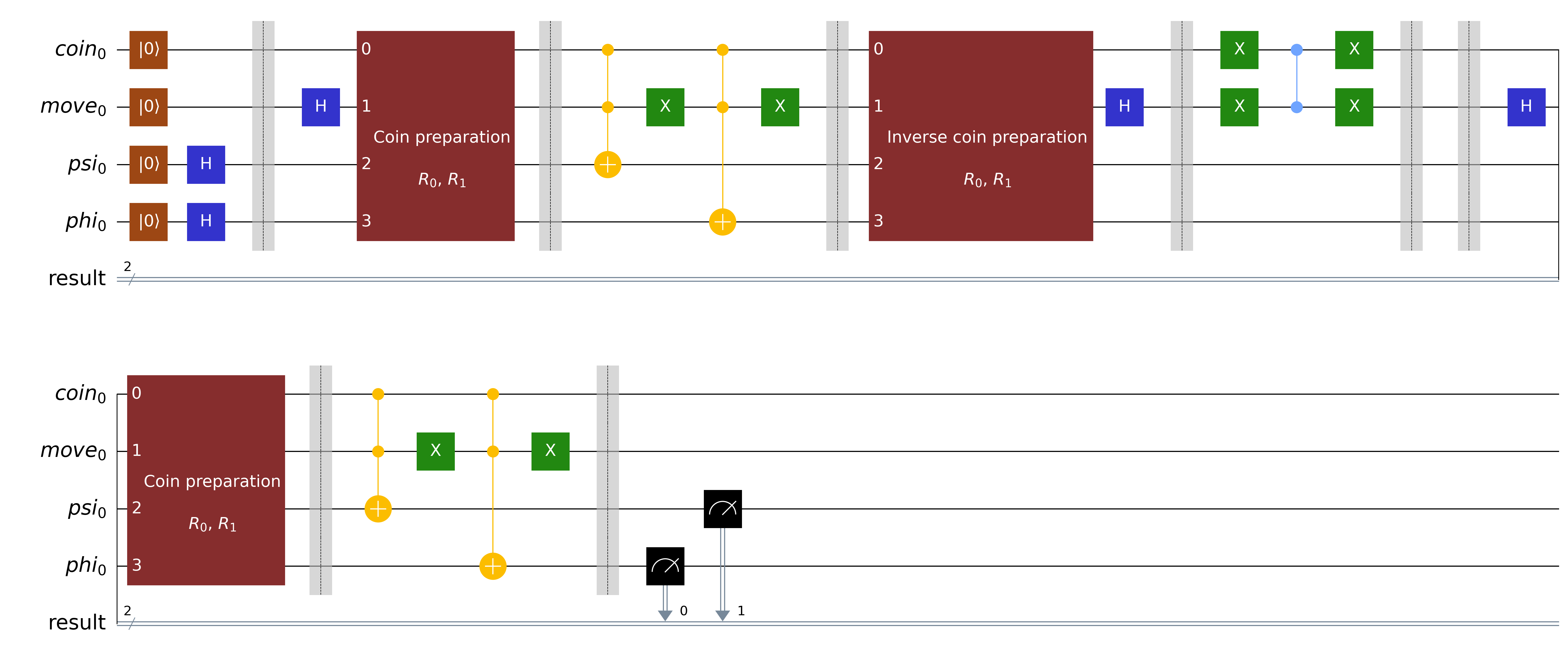}
\caption{Implementation of the full quantum Metropolis circuit from section \ref{sec:Hardware_experiment_results} implemented in actual quantum hardware, using the coin flip rotation described in algorithm \ref{alg:coin_flip}. The steps of the circuit are separated by barriers for a better identification (from left to right, and from top to bottom): first put $\phi$ and $\psi$ in superposition. Then, for each of the two steps $\tilde{W}$ from \eqref{quantum walk operator tilde W}: put the move register (that controls which angle to move) in a superposition, operator $V$, and prepare the coin, $B$. Then, controlled on the coin being in state $\ket{1}$ and the move register the corresponding angle, change the value of $\psi$ (if move $= \ket{1}$) or $\phi$ (if move $= \ket{0}$), denoted by $F$. Then we uncompute the coin, $B^\dagger$ and the move preparation, $V^\dagger$, and perform the phase flip on $\ket{\text{move}}\ket{\text{coin}}= \ket{00}$ represented by $R$ in eq. \eqref{quantum walk operator tilde W}, and by the four X gates sandwitching a C-Z in the circuit. The second quantum walk step proceeds equally (with different value of $\beta$ and the rotations), but now we do not have to uncompute the move and coin registers before measuring $\phi$ and $\psi$ because it is the last step.}
\label{fig:circuit}\end{figure*}

Lastly, large differences in the exponents exist depending on the initialization criteria used, although the same argument on why this is the case given in section \ref{sec:Fixed_beta_results} applies here too. 

\section{\label{sec:Hardware_experiment_results} Experiments in IBMQ Casablanca}

\subsection{\label{sec:Experiment_conditions} Experimental setup}

We have also performed experiments in the IBMQ Casablanca system. In contrast to previous simulations, due to the low signal to noise ratio in the available quantum hardware, it does not make much sense to directly compare the values of the TTS figure of merit. Instead, the objective here is to show that we can implement a two-step quantum walk (the minimum required to produce interference) and still see an increase in probability associated with the correct state. Since we are heavily constrained in the depth of the quantum circuit we can implement, we experiment only with dipeptides, and with 1 bit of precision in the rotation angles: that is $\phi$ and $\psi$ can be either $0$ or $\pi$. 

In this quantum circuit, depicted in figure \ref{fig:circuit} together with algorithm \ref{alg:coin_flip}, we will have 4 qubits, two encoding the value of the angles $\ket{\cdot}_{\phi}$, $\ket{\cdot}_{\psi}$, a coin qubit $\ket{\cdot}_C$, and the final one indicating what angle register to update in the next step, $\ket{\cdot}_M$. Additionally, we always start from the uniform superposition $\ket{+}_{\phi}\ket{+}_{\psi}$. We then perform 2 quantum walk steps $\tilde{W}$ with values of $\beta$ empirically chosen $0.1$ and $1$ to have large probabilities of measuring $\ket{0}_{\phi}\ket{0}_{\psi}$, where we encode the ground state. 

\begin{algorithm}
\caption{Coin preparation subroutine for the hardware-adapted circuit.}\label{alg:coin_flip}
\begin{algorithmic}[1]
\Procedure{Classically}{}
\State Compute rotation angles starting from $\ket{1}_C$.
\State angles($\ket{0}_{\phi}\ket{0}_{\psi}\ket{0}_{M}$) $\approx$ angles($\ket{0}_{\phi}\ket{1}_{\psi}\ket{0}_{M}$), so take their average in $R_0$.
\State angles($\ket{0}_{\phi}\ket{0}_{\psi}\ket{1}_{M}$) $\approx$ angles($\ket{1}_{\phi}\ket{0}_{\psi}\ket{1}_{M}$), so take their average in $R_1$.
\EndProcedure
\Procedure{In the quantum circuit}{}
\State Apply $X$-gate to $\ket{\cdot}_C$.
\State Controlled on $\ket{0}_{\phi}\ket{0}_{M}$, X-rotate $\ket{\cdot}_C$ by $R_0$.
\State Controlled on $\ket{0}_{\psi}\ket{1}_{M}$, X-rotate $\ket{\cdot}_C$ by $R_1$.
\EndProcedure
\end{algorithmic}
\end{algorithm}

We depict the circuit in figure \ref{fig:circuit}, with the coin preparation subroutine defined in algorithm \ref{alg:coin_flip}. The latter is the most costly part of the quantum circuit. This hardware-adapted coin flip contains two main simplifications in order to minimize the length of the circuit as much as possible. First, since half the possible moves are accepted with probability 1, we apply an $X$-gate on coin $X\ket{0}_C = \ket{1}_C$, and rotate back to $\ket{0}_C$ from there when needed. This halves the number of necessary control rotations.

Second, we group similar angle values together, reducing their number and also the number of control qubits. For example, one should in principle perform a controlled rotation on the different values of $\ket{\cdot}_{\phi}\ket{\cdot}_{\psi}\ket{\cdot}_{M}$. However, in our cases of interest the angle corresponding to $\ket{0}_{\phi}\ket{0}_{\psi}\ket{1}_{M}$ is close to the rotation angle that should be controlled by $\ket{1}_{\phi}\ket{0}_{\psi}\ket{1}_{M}$. For such reason, we perform a controlled rotation controlled on $\ket{0}_{\psi}\ket{1}_{M}$, with the average of those two angles, as can be seen in algorithm \ref{alg:coin_flip}. This second approximation still makes the state $\ket{00}$ to be the lowest energy state and the most likely in the noiseless simulation of the error, as can be seen in figure \ref{fig:Hardware}.

There is one more important precision to be made: since our implementation of the quantum circuit in the IBMQ Casablanca system has 176 basic gates of depth even after being heavily optimized by \textit{Qiskit} transpiler, we need a way to tell whether what we are measuring is only noise or relevant information survives it. Our first attempt to distinguish these two cases was to use the natural technique of zero-noise extrapolation, where additional gates that do not change the theoretical expected value of the circuit are added, but introduce additional noise \cite{temme2017error}. By measuring how the measured probabilities change, one can extrapolate backwards to the theoretical `zero noise' case. Unfortunately, the depth of the circuit is already so large that it does not work: it does not converge or else returns unrealistic results, at least when attempted with the software library \textit{Mitiq} \cite{larose2020mitiq}. 

For this reason we need to find a way out that will only be valid because our circuit is parameterised in terms of the angles and the values of $\beta$. If we were to set the value of $\beta$ to $0$, the theoretical result would be $1/4$ as there are 4 possible states and each step would always be accepted. As a consequence, our strategy consists of trying to detect changes in the probability when we use $\beta(\bm{t}) = (0,0)$ or $\beta(\bm{t}) =(0.1,1)$. The notation $\beta(\bm{t})$ denotes the value of $\beta$ chosen at each of those two steps.

For the experiment, we reserved 3 hours of usage of the IBMQ Casablanca system, with quantum volume 32. During that time we run a total of 163840 circuits for each of the 8 randomly chosen dipeptides, and 204800 for $\beta(\bm{t}) = (0,0)$ as a baseline. 

\subsection{\label{sec:Experiment_results} Results}

\begin{figure}[t]
    \centering
    \includegraphics[width = 80mm]{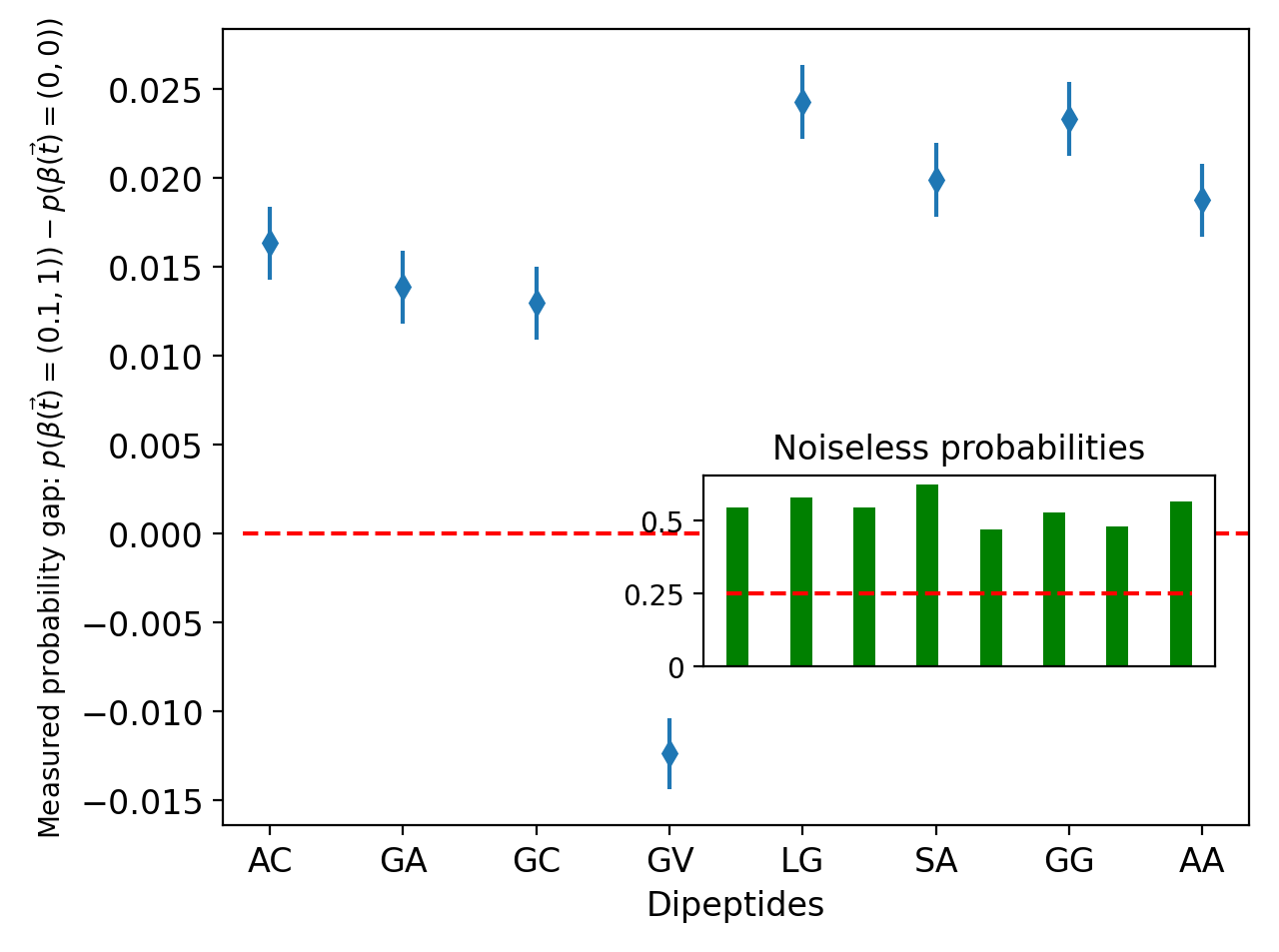}
    \caption{Results from hardware measurements corresponding to the experimental realization of the quantum Metropolis described in section \ref{sec:Hardware_experiment_results} and the circuit depicted in figure \ref{fig:circuit}. For each dipeptide we perform a student t-test to check whether the average success probabilities for $\beta(\bm{t}) = (0,0)$ and $\beta(\bm{t}) = (0.1,1)$ are actually different \cite{student1908ttest}. The largest p-value measured in all 8 cases is $3.94\cdot 10^{-18}$, indicating that in all cases the difference is significant. For each dipeptide we run 163840 times the circuit, and for the baseline 204800 times. The noiseless probabilities also contain the approximations we described in algorithm \ref{alg:coin_flip}.}
    \label{fig:Hardware}
\end{figure}

We depict the results from our experiments in figure \ref{fig:Hardware}. The inner figure represents the theoretical increase in probability when using $\beta(\textbf{t}) = (0.1,1)$ vs using $\beta(\textbf{t}) = (0,0)$. In the latter case the probability of measuring $\ket{00}$ is $1/4$, while the former probability is around $1/2$.

We want to contrast that theoretical probability increase with what we are able to measure in the hardware, depicted in the the outer plot. In 7 of the 8 cases we are able to measure such increase. The outlier, glycylvaline, is surprising because is the dipeptide that in the simulation shows the greatest theoretical probability of measuring $\ket{00}$. We can only hypothesise that this is due to some experimental imperfection.

We carry out some statistical analysis of the results depicted in the figure. The measurements for $\beta(\textbf{t}) = (0,0)$ and $\beta(\textbf{t}) = (0.1,1)$ are modelled as binomial distributions. That is, in the first case there will be a probability $p_0$ of measuring $\ket{00}$, while in the second case the probability will be $p_1$. The question we want to answer is: is $p_1 \neq p_0$? If such were the case, the probability difference would be attributable to the value of $\beta$, as the circuit is otherwise the same.

For each dipeptide we then run a t-test \cite{student1908ttest} on the measurements of each dipeptide with $\beta(\textbf{t}) = (0.1,1)$ against the measurements for the $\beta(\textbf{t}) = (0,0)$ case, finding out that the largest p-value is $3.94\cdot 10^{-18}$. This indicates that $p_1 \neq p_0$ with high probability in all cases.

Moreover, we can also give the standard deviation for the best estimator of $p_0$ and $p_1$, the measured averages $\hat{p}_0$ and $\hat{p}_1$. Remember that for a sample from a binomial distribution with probability $p$ bounded away from 0 and 1 and $n$ large, $X \sim Bin(n,p)$, \cite{brown2001interval}
\begin{equation}
    \mathbb{E}[X] = np, \qquad Var[X] = np(1-p).
\end{equation}
This means that the standard deviation can be estimated as 
\begin{equation}
    \hat{\sigma} = \sqrt{\frac{\hat{p}(1-\hat{p})}{n}},
\end{equation}
where $\hat{p}$ is the estimated probability. We depict as error bars in figure \ref{fig:Hardware} the sum of such estimators for the case of $\beta(\textbf{t}) = (0,0)$ and $\beta(\textbf{t}) = (0.1,1)$. Furthermore, based on the normal approximation to the binomial distribution, the interval confidence given by $\hat{\sigma}$ for either $\beta(\textbf{t}) = (0,0)$ or $\beta(\textbf{t}) = (0.1,1)$ might be interpreted as the $68\%$ confidence interval for the true value of $p$ \cite{brown2001interval, wallis2013binomial}. $\hat{p}\pm 2\hat{\sigma}$ can be similarly interpreted as the $95\%$ confidence interval, which would correspond to roughly duplicating the error bars.

\section{\label{sec:Conclusions} Conclusions and outlook}

We have studied how quantum computing might complement modern machine learning techniques to predict the structure of proteins. For that, we have introduced QFold, an algorithm that implements a quantum Metropolis algorithm using as a starting guess the output of a machine learning algorithm. In our case this will be a simplified implementation of AlphaFold algorithm named Minifold, which could be substituted by any initialization module that uses future improvement of such deep learning techniques.

An important feature of QFold is that it is a realistic description of the protein structure, meaning that the description of the folded structure relies on the actual torsion angles that describe the final conformation of the protein. This is realised by the number of bits $b$ used to fix the precision of those angles, for which a moderate value of $b = 5$ or $b = 6$ would be as accurate as the original AlphaFold. This is in sharp contrast with the rigid lattice models used to approximate protein folding in the majority of quantum computing algorithmic proposals for protein structure prediction.  Although in our current simulations presented in this work the range of the precision is limited by the resources of the classical simulation, nothing prevents QFold from reaching a realistic accurate precision once a fully fledged quantum computer is at our disposal, since our formulation is fully scalable within a fault-tolerant scenario.

There are also many other details that we have only scarcely commented, such as side-chain fixation or interaction of the protein with its environment. However, the reader must take into account that our current implementation and simulation is only a proof-of-concept which, perhaps more importantly, can be scaled up to include all those details. To do so, one only needs to choose an appropriate energy evaluation function to replace our oracle, such as for example Rosetta scoring function. Then, one must convert it to a reversible function (which can always be done with small overhead) and substitute the classical gates by their quantum counterparts. Additionally, side-chain dihedral angles can be handled with the same algorithm that we have used, with the corresponding increased complexity. In conclusion, QFold represents perhaps the first quantum algorithm able to incorporate all reality details common in classical algorithms, and the first to measure hints of quantum advantage in protein structure prediction.

The quantum Metropolis algorithm itself relies on the construction of a coined version of Szegedy quantum walk \cite{lemieux2019efficient}, and we use the Total Time to Solution defined in \eqref{TTS}, as a figure of merit. Our construction of this quantum Metropolis algorithm represents the first main contribution of our work.

The second main contribution is an analysis of the expected quantum advantage in protein structure prediction, that although moderate in the scaling exponent with the Minifold initialization, could represent a large difference in the expected physical time needed to find the minimal energy configuration in proteins of average size due to the exponential nature of this combinatorial problem. This quantum advantage analysis is also performed for different realistic annealing schedules, indicating that the out-of-equilibrium quantum Metropolis algorithms can show a similar quantum advantage, and can even improve the advantage for the fixed beta case, as can be seen from table \ref{table:exponents}. However, it is also worth noticing that we do not expect such quantum advantage to be be profitable with early fault-tolerant quantum computers due to the error correction overhead \cite{babbush2021focus}. As such we acknowledge the need to find ways to improve both the error correction overhead and the quantum advantage. The third contribution is a minimal implementation of our algorithm in actual quantum software.

Our results for the computation of protein structure prediction provide further support to the development of classical simulations of quantum hardware. A clear message from our simulations is that it is worthwhile developing quantum software and classical simulations of ideal quantum computers in order to confirm that certain quantum algorithms provide a realistic quantum advantage. Some of the quantum algorithms that must be assessed in such conditions are those based on quantum walks like quantum Metropolis variants with fast annealing schedule. For them, the complexity scales as $O(\delta^{-a})$, $\delta$ the eigenvalue gap, and $a>0$ dependent on the specific application and realization of the quantum Metropolis. This is in contrast with \textit{standard quantum walks}, where the classical complexity scales as $O(\delta^{-1})$ and the quantum complexity as $O(\delta^{-1/2})$, as can be seen in appendix \ref{sec:OOE-Metropolis_algorithm}. However, it would be naïve to consider this quadratic quantum advantage as an achievable and useful one in problems similar to ours. Instead, one should aim to compare quantum algorithms with the classical ones used in practice, where heuristic annealing schedules are used instead. As a consequence, finding out the precise values of the corresponding exponent is of great importance since it is a measure of the quantum advantage. 
For this reason, not all efforts should be devoted to finding a quantum advantage solely with NISQ devices, but also to continue investigating the real possibilities of universal quantum computers when they become available.


We would also like to point out that despite the great advances achieved by the new classical methods in the latest editions of the CASP competition \cite{CASP}, there is still considerable room for improvement and there are many gaps in the problem of protein folding that await to be filled. These include understanding protein-protein and protein-ligand interactions, real time folding evolution to the equilibrium configuration, the dark proteome, proteins with dynamical configurations like the intrinsically disordered proteins (IDP) and so on and so forth. Crucially, we believe that the current limitations of the training data sets, which are biased towards easily to be crystallized proteins, puts constraints on what can be achieved using only deep learning techniques. Our present research is an attempt to explore techniques that address these limitations.


Finally, future work should search for heuristic variants of the quantum metropolis algorithms \cite{lemieux2019efficient}, especially those offering larger polynomial advantages than found in our simulations.
Those quantum Metropolis algorithms could be used in a variety of domains, and as such a detailed analysis, both theoretical and experimental, of the expected quantum advantage in each case seems desirable.

\section{Acknowledgements}
P.A.M.C and R.C contributed equally to this work. We would like to thank kind advice from Jaime Sevilla on von Mises distribution and statistical t-tests, Alvaro Mart\'inez del Pozo and Antonio Rey on protein folding, Andrew W. Senior on minor details of his AlphaFold article, Carmen Recio, Juan G\'omez, Juan Cruz Benito, Kevin Krsulich and Maddy Todd on the usage of Qiskit, and Jessica Lemieux and the late David Poulin on aspects of the quantum Metropolis algorithm. We acknowledge the access to advanced services provided by the IBM Quantum Researchers Program. We also thank Quasar Science for facilitating the access to the AWS resources.
We acknowledge financial support from the Spanish MINECO grants MINECO/FEDER Projects FIS 2017-91460-EXP, PGC2018-099169-B-I00 FIS-2018 and from CAM/FEDER Project No. S2018/TCS-4342 (QUITEMAD-CM). The research of M.A.M.-D. has been partially supported by the U.S. Army Research Office through Grant No.  W911NF-14-1-0103. P. A. M. C. thanks  the support of a MECD grant FPU17/03620, and R.C. the support of a CAM grant IND2019/TIC17146. 

\appendix

\section{\label{sec:Szegedy} Szegedy quantum walks}

In order to explain what are quantum walks, we need to introduce Markov Chain. Given a configuration space $\Omega$, a Markov Chain is a stochastic model over $\Omega$, with transition matrix $\mathcal{W}_{ij}$, that specifies the probability of transition that does not depend on previous states but only on the present one. Random walks are the process of moving across $\Omega$ according to $\mathcal{W}_{ij}$.

Quantum walks are the quantum equivalent of random walks \cite{portugal2013quantum}. The most famous and used quantum walks are those defined by Ambainis \cite{ambainis2007quantum} and Szegedy \cite{szegedy2004quantum}, although several posterior generalisations and improvements have been developed such as those in \cite{magniez2011search}. Quantum walks often achieve a quadratic advantage in the hitting time of a target state with respect to the spectral gap, defined below, and are widely used in several other algorithms \cite{paparo2012google,paparo2013quantum,paparo2014quantum}, as we shall see. 

Szegedy quantum walks are not usually defined using a coin, but rather on a bipartite walk. This implies duplicating the Hilbert space, and defining the unitary
\begin{subequations}
\begin{equation}
    U\ket{j}\ket{0} := \ket{j} \sum_{i \in \Omega} \sqrt{\mathcal{W}_{ji}}\ket{i} = \ket{j}\ket{p_j}
\end{equation}
and also the closely related
\begin{equation}
    V\ket{0}\ket{i} := \sum_{j \in \Omega}\sqrt{\mathcal{W}_{ij}} \ket{j}  \ket{i} = \ket{p_i}\ket{i}.
\end{equation}
\end{subequations}

\begin{figure*}[t]
\centering
\includegraphics[width=0.75\textwidth]{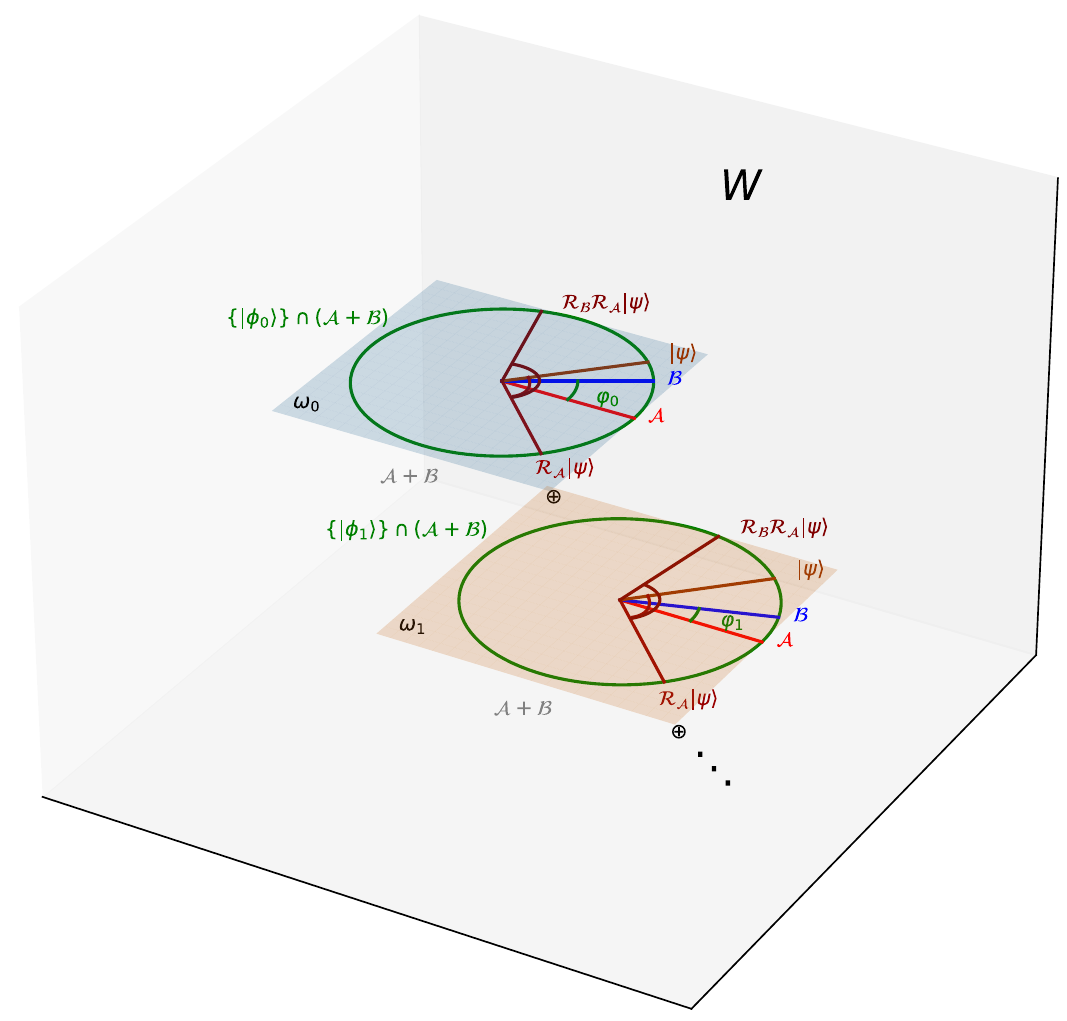}
\caption{Geometrical visualization of a quantum walk operator $W$ of Szegedy type. $W$ performs a series of rotations that in the subspace $\mathcal{A} + \mathcal{B}$, defined in \eqref{Subspaces A and B} with their corresponding rotation operators \eqref{subespace rotations}, may be written as a block diagonal  matrix, where each block is a 2-dimensional rotation $\omega_j = R(2\varphi_j)$ given by \eqref{block rotation matrix in A+B}. This figure represents the direct sum of Grover-like rotations in the subspace spanned by $\mathcal{A}+\mathcal{B}$, and therefore $W$. This quantum walk operator represents equation \eqref{W definition} from section \ref{sec: Quantum Metropolis}.}
\label{fig:quantum_walk}
\end{figure*}

$\mathcal{W}_{ij}$ must be understood as the probability that state $\ket{i}$ transitions to state $\ket{j}$.
We can check that these operators fulfil that $SU = VS $, for $S$ the Swap operation between the first and second Hilbert subspace. The cost of applying $U$ is usually called (quantum) update cost. Define also the subspaces
\begin{subequations}
\begin{equation}
    \mathcal{A}:= \text{span} \{\ket{j}\ket{0}: j\in \Omega\}
    \label{A subspace}
\end{equation}
and
\begin{equation}
    \mathcal{B}:= U^\dagger S U \mathcal{A}  = U^\dagger V S \mathcal{A}.
    \label{B subspace}
\end{equation}
\label{Subspaces A and B}
\end{subequations}
Having defined $U$ and $V$, we can define $M:= U^\dagger V S$, a matrix with entries valued $\braket{i,0|U^\dagger V S|j,0}= \sqrt{\mathcal{W}_{ji}}\sqrt{\mathcal{W}_{ij}} = \sqrt{\pi_i/\pi_j}\mathcal{W}_{ij}$, the last equality thanks to the detailed balance equation \footnote{Notice that in most texts the definition of $M$ does not explicitly include $S$. It is assumed implicitly though.}. In fact, in matrix terms it is usually written $M = D_\pi^{-1/2}\mathcal{W} D_\pi^{1/2}$ where we have written $D_\pi$ to indicate the diagonal matrix with the entries of the equilibrium state-vector of probabilities, $\pi$. This implies that $\mathcal{W}$ and $M$ have the same spectrum $1 > \lambda_0 \geq  ... \geq \lambda_{d-2}\geq  0$, as the matrix $\mathcal{W}$ is positive definite, $p^T\mathcal{W}p \in [0,1]$, and of size $d$. The corresponding eigenstates are $\ket{\phi_j}\ket{0}$, and phases $\varphi_j = \arccos{\lambda_j}$. In particular $\ket{\phi}=\sum_i \sqrt{\pi_i} \ket{i}$, is the equilibrium distribution.

We can also define the projectors around $\mathcal{A}$ and $\mathcal{B}$ as $\Pi_{\mathcal{A}}$ and $\Pi_{\mathcal{B}}$
\begin{subequations}
\begin{equation}
    \Pi_{\mathcal{A}}:= (\mathbf{1}\otimes \ket{0}\bra{0}),
    \label{A subspace proyector}
\end{equation}
\begin{equation}
    \Pi_{\mathcal{B}}:= U^\dagger V S (\mathbf{1}\otimes \ket{0}\bra{0}) S V^\dagger U
    \label{B subspace proyector}
\end{equation}
\end{subequations}
with their corresponding rotations 
\begin{subequations}
\begin{equation}
R_\mathcal{A} = 2\Pi_\mathcal{A} - \mathbf{1},
\end{equation}
\begin{equation}
R_\mathcal{B} = 2\Pi_\mathcal{B} - \mathbf{1}.
\end{equation}
\label{subespace rotations}
\end{subequations}
Using this rotation we further define a quantum walk step as we did in the main text equation \eqref{W definition},
\begin{equation}
 W= R_\mathcal{B}R_\mathcal{A}= U^\dagger S U R_\mathcal{A} U^\dagger S U R_\mathcal{A}. \label{W definition appendix}
\end{equation}
Using the previous expressions, we can state, using $\Pi_\mathcal{A}\ket{\phi_j}\ket{0}=\ket{\phi_j}\ket{0}$
\begin{subequations}
    \begin{equation}
        \Pi_\mathcal{A}U^\dagger VS \ket{\phi_j}\ket{0} = \cos \phi_j \ket{\phi_j}\ket{0},
    \end{equation}
    and
    \begin{equation}
        \Pi_\mathcal{B} \ket{\phi_j}\ket{0} =  U^\dagger V S \cos \phi_j \ket{\phi_j}\ket{0}.
    \label{effect of B proyector}
    \end{equation}
\end{subequations}
\eqref{effect of B proyector} is true (supplementary material \cite{yung2012quantum}), because
\begin{equation}
    \Pi_\mathcal{A} U^\dagger V S \Pi_\mathcal{A} = \Pi_\mathcal{A} SV^\dagger U \Pi_\mathcal{A}
\end{equation}
due to 
\begin{equation}
\begin{split}
    &\braket{\phi_j,0 |\Pi_\mathcal{A} U^\dagger V S \Pi_\mathcal{A}|\phi_j,0} = \lambda_j\\
    &= \lambda_j^\dagger =\braket{\phi_j,0 |\Pi_\mathcal{A} SV^\dagger U \Pi_\mathcal{A}|\phi_j,0}.
\end{split}
\end{equation}
Thus $W$ will preserve the subspace spanned by $\{\ket{\phi_j}\ket{0}, U^\dagger V S \ket{\phi_j}\ket{0} \}$, which is invariant under $\Pi_{\mathcal{A}}$ and $\Pi_{\mathcal{B}}$; mirroring the situation in the Grover algorithm \cite{Grover}.
Also, as a consequence of the previous, and of the fact that in $\mathcal{A}+\mathcal{B}$ operator $W$ has eigenvalues $e^{2i\varphi_j}$ \cite{magniez2011search,szegedy2004quantum}, in such subspace the operator $W$ can be written as a block diagonal operator with matrices
\begin{equation}
w_j =
    \begin{pmatrix}
    \cos(2\varphi_j) & -\sin(2\varphi_j)\\
    \sin(2\varphi_j) &
    \cos(2\varphi_j)
    \end{pmatrix}.
    \label{block rotation matrix in A+B}
\end{equation}
The rotation in the subspace corresponding to eigenvalue $1$ will be an identity matrix.

Finally, notice that the eigenvalue gap of $\mathcal{W}$ is defined as $\delta = 1 - \lambda_0$, and in general the hitting time of a classical walk will grow like $O(\delta^{-1})$ (Proposition 1 of \cite{magniez2011search}). On the other hand, the hitting time of the quantum walk will scale like $O(\Delta^{-1})$ (Theorem 6 of \cite{magniez2011search}), where $\Delta := 2\varphi_0$. However, $\Delta \geq 2\sqrt{1-|\lambda_0|^2}\geq 2\sqrt{\delta}$, so $\Delta = \Omega(\delta^{1/2})$. In fact, writing $\delta = 1- \lambda_0 = 1 - \cos \varphi_0$, and expanding in Taylor series $\cos \varphi_0 = 1- \frac{\varphi_0^2}{2} + \frac{\varphi_0^4}{24} + O(\varphi_0^6)$, we can see that 
\begin{equation}
    \frac{\varphi_0^2}{2} \geq 1 -\cos\varphi_0 \geq \frac{\varphi_0^2}{2} - \frac{\varphi_0^4}{24}.
\end{equation}
Using the definitions of $\delta$ and $\Delta$ and the fact that $\varphi_0 \in (0, \pi/2)$, it is immediate
\begin{equation}
\begin{split}
    \frac{\Delta^2}{8}\geq \delta \geq \frac{\Delta^2}{8} - \frac{\Delta^4}{2^4\cdot 24} &= \frac{\Delta^2}{8}\left(1 - \frac{\Delta^2}{ 2\cdot 24}\right)\\
    &\geq \frac{\Delta^2}{8}\left(1 - \frac{\pi^2}{2\cdot 24}\right).
\end{split}
\end{equation}
Consequently, $\Delta = \Theta(\delta^{1/2})$, and this is the reason quantum walks are quadratically faster than their classical counterparts.

\section{\label{sec:OOE-Metropolis_algorithm} Mathematical description of the out-of-equilibrium quantum Metropolis algorithm}

In appendix \ref{sec:Szegedy} we have reviewed the Szegedy quantum walk. In this appendix, we present a quantum Metropolis-Hasting algorithm based on the use of Szegedy walks.

The objective of the Metropolis-Hastings algorithm is sampling from the Gibbs distribution $\rho^\beta = \ket{\pi^\beta}\bra{\pi^\beta} =  Z^{-1}(\beta)\sum_{\bm{\phi}\in \Omega} e^{-\beta E(\bm{\phi})}\ket{\bm{\phi}}\bra{\bm{\phi}}$, where $E(\bm{\phi})$ is the energy of a given configuration of angles of the molecule, $\beta$ plays the role of the inverse of a temperature that will be lowered during the process, and $Z(\beta) =\sum_{\bm{\phi}\in \Omega} e^{-\beta E(\bm{\phi})}$ a normalization factor. $\Omega$ represents the configuration space, in our case the possible values the torsion angles may take. One can immediately notice that if $\beta$ is sufficiently large, only the configuration with the lowest possible energy will appear when sampling from that state with high probability. Thus, we wish to prepare one such state to be able to find the configuration with the lowest energy, in our case the folded state of the protein in nature.

One way to construct such a distribution $\pi^\beta$ is to create a rapidly mixing Markov Chain that has as equilibrium distribution $\pi^\beta$. Such Markov Chain is characterized, at a given $\beta$, by a transition matrix $\mathcal{W}$ that induces a random walk over the possible states. That is, $\mathcal{W}$ maps a given distribution $p$ to another $p'= \mathcal{W}p$. Let us introduce some useful definitions: an aperiodic walk is called \textit{irreducible} if any state in $\Omega$ can be accessed from any other state in $\Omega$, although not necessarily in a single step. Additionally, we will say that a walk is \textit{reversible} when it fulfills the detailed balance condition
\begin{equation}
    \mathcal{W}_{j,i}^{\beta}\pi^{\beta}_{i} = \mathcal{W}_{i,j}^{\beta}\pi^{\beta}_{j}. \label{detailed balance}
\end{equation}

Metropolis-Hastings algorithm uses the following transition matrix
\begin{equation}
  \mathcal{W}_{ij}=\begin{cases}
    T_{ij}A_{ij}, & \text{if $i\neq j$}\\
    1-\sum_{k\neq j} T_{kj}A_{kj}, & \text{if $i= j$},
  \end{cases}
\end{equation}
where, as given in the main text equation \eqref{Metropolis update probability},
\begin{equation}
    A_{ij}= \min \left(1,e^{-\beta(E_i-E_j)}\right),
    \label{Metropolis update probability appendix}
\end{equation}
and
\begin{equation}
  T_{ij}=\begin{cases}
    \frac{1}{N} & \text{there is a move connecting $j$ to $i$}\\
    0, & \text{else},
  \end{cases}
\end{equation}
for $N$ the number of possible outgoing movements from state $j$, which we assume to be independent of the current state, as is the case for our particular problem. In the case of the Metropolis-Hastings algorithm, the detailed balance condition is fulfilled with the above definitions.

Having defined the Metropolis algorithm, we now want to quantize it using the previous section.
There have been several proposals to quantize the Metropolis algorithm, as can be seen in Appendix \ref{sec:OOE-Metropolis_algorithm}. They often rely on slow evolution from $\ket{\pi_t}\rightarrow \ket{\pi_{t+1}}$ using variations of Amplitude Amplification, until a large $\beta$ has been achieved.

However, most often this Metropolis algorithms are used outside of equilibrium, something that is not often worked out in the previous approaches. There are at least two ways to perform the quantization of the out-of-equilibrium Metropolis-Hastings algorithm \cite{lemieux2019efficient}. The first one, called `Zeno with rewind' \cite{temme2011quantum} uses a simpler version of previous quantum walks, where instead of amplitude amplification, quantum phase estimation is used on the operators $W_j$, so that measuring the first eigenvalue means that we have prepared the corresponding eigenvector $\ket{\psi_t^0} = \ket{\pi_t}$ \cite{somma2007quantum}. Since the eigenvalue gap is $\Delta_t$, the cost of quantum phase estimation is $O(\Delta_t^{-1})$.

Define measurements $Q_t = \ket{\pi_t}\bra{\pi_t}$ and $Q^\perp_t = \mathbf{1}-\ket{\pi_t}\bra{\pi_t}$, for $t$ an index indicating the step of the cooling schedule. Performing these measurements consists, as we have mentioned, in performing phase estimation of the corresponding operator $\mathcal{W}_{t}$, at cost $\Delta_t^{-1}$. If a measurement of type $Q^\perp_t$ indicates a restart of the whole algorithm it is called `without rewind'. However, if measurement $Q^\perp_t$ is obtained, one can perform phase estimation of $\mathcal{W}_{t-1}$. If $|\braket{\pi_t|\pi_{t-1}}|^2 = F_t^2$, then the transition probability between $Q_j \leftrightarrow Q_{t-1}$ and between $Q^\perp_t \leftrightarrow Q^\perp_{t-1}$ is given by $F_t^2$; and the transition probability between $Q_t \leftrightarrow Q^\perp_{t-1}$ and between $Q^\perp_t\leftrightarrow Q_{t-1}$ is given by $1-F_t^2$, so in a logarithmic number of steps one can recover state $\ket{\pi_{t-1}}$ or $\ket{\pi_{t}}$.

The second proposal is to perform the unitary heuristic procedure \cite{lemieux2019efficient}
\begin{equation}
    \ket{\psi(L)} = W_L ... W_1 \ket{\pi_0}.
\end{equation}
This is in some ways the simplest way one would think of quantizing the Metropolis-Hastings algorithm, implementing a quantum walk instead of a random walk. It is very similar to the classical way of performing many steps of the classical walk, slowly increasing $\beta$ until one arrives to the aim temperature. In addition, this procedure is significantly simpler than the previously explained ones because it does not require to perform phase estimation on $W_t$.

Two more innovations are introduced by \cite{lemieux2019efficient}. In the first place, a heuristic Total Time to Solution (TTS) is defined. Assuming some start distribution, if operators $W$ are applied $t$ times, the probability of success is given by $p(t)$. In order to be successful with constant probability $1-\delta$ that means repeating the procedure $\log(1-\delta)/\log(1-p(t))$ times. The total expected time to success is then
\begin{equation}
    TTS(t):= t \frac{\log (1-\delta)}{\log ( 1-p(t))},
\end{equation}
as also indicated in the main text equation \eqref{TTS}.

Additionally \cite{lemieux2019efficient} constructed a coined version of Szegedy operator, called $\tilde{W}$ such that the quantum walk operator operates on three registers $\ket{x}_S\ket{z}_M\ket{b}_C$. $S$ stands for the codification of the state, $M$ for the codification of possible movements, and $C$ the Boltzmann coin. Their operator $\tilde{W}$ is equivalent to the Szegedy operator under a conjugation by operator $Y$ that maps states in the second register of Szegedy's original quantum walk to the moves that are needed to achieve them \cite{lemieux2019efficient}:
\begin{subequations}
\begin{equation}
    Y^\dagger (R_{\mathcal{A}}U^\dagger S U) Y = \tilde{W}
\end{equation}
with
\begin{equation}
    Y: \ket{x}_{S}\ket{y}_{S'} \mapsto \ket{x}_S\ket{m}_M \text{ such that } m(x) = y.
\end{equation}
\end{subequations}
The modified quantum walk is
\begin{equation}
    \tilde{W} = R V^\dagger B^\dagger F B V,
\end{equation}
with
\begin{subequations}
\begin{equation}
    V: \ket{0}_M \rightarrow \sum_m N^{-1/2} \ket{m}_M,
\end{equation}
\begin{equation}
\begin{split}
    B: &\ket{x}_S\ket{m}_M\ket{0}_C \rightarrow\\
    &\ket{x}_S\ket{m}_M\left(\sqrt{1-A_{x\cdot z_j,x}}\ket{0}_C+\sqrt{A_{x\cdot z_j,x}}\ket{1}_C\right), \label{coin flip}
\end{split}
\end{equation}
\begin{equation}
\begin{split}
    F: &\ket{x}_S\ket{m}_M\ket{b}_C \rightarrow \\
    &\ket{b\cdot m(x) + (1-b)\cdot x}_S\ket{m}_M\ket{b}_C
\end{split}
\end{equation}
and
\begin{equation}
\begin{split}
    R: &\ket{0}_M\ket{0}_C \rightarrow - \ket{0}_M\ket{0}_C\\
    & \ket{m}_M\ket{b}_C \rightarrow \ket{m}_M\ket{b}_C, \quad (m,b)\neq (0,0)
\end{split}
\end{equation}
\end{subequations}
Here $V$ proposes different moves, $B$ prepares the Boltzmann coin, $F$ flips the bits necessary to prepare the new state, conditional on the Boltzmann coin being in state 1, and $R$ is a reflection operator on state $(0,0)$ for the coin and movement registers. Although our encoding of the operators and states is slightly different, the algorithm that we have used is this one, mainly due to its simplicity.
In conclusion, the TTS(L) is calculated as (see eq. 35 \cite{lemieux2019efficient})
\begin{equation}
    TTS(L)= L \frac{\log (1-\delta)}{\log ( 1-|\braket{\pi^g|\tilde{W}_L...\tilde{W}_1|\pi_0}|^2)}
    \label{quantum TTS definition}
\end{equation}
with $\pi^g$ the ground state of the Hamiltonian.

\section{Circuit implementation of the coin flip subroutine in IBMQ}

\begin{figure*}[h!]
\centering
\includegraphics[width=\textwidth]{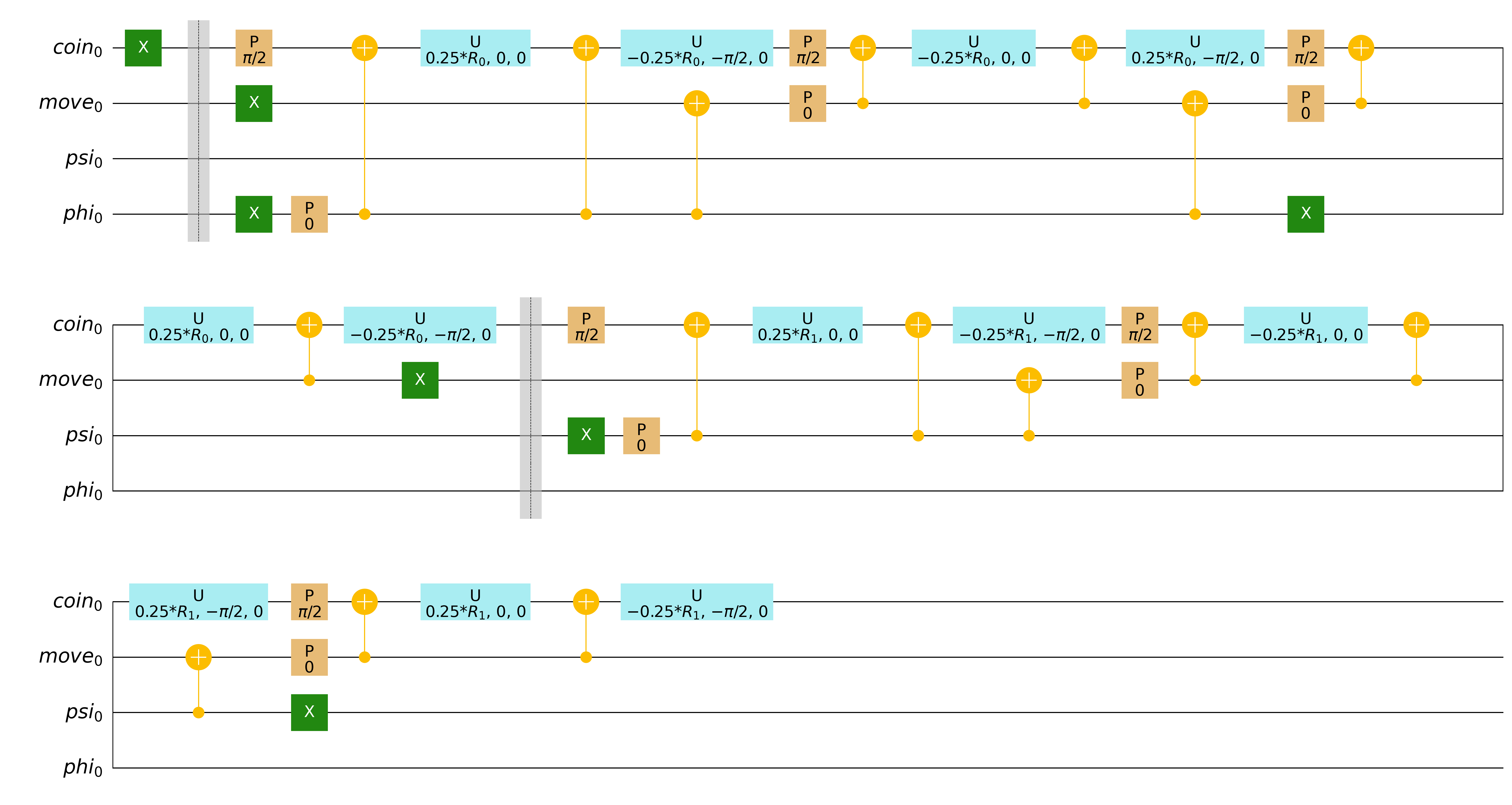}
\caption{The circuit has two key simplifications that reduce the depth. The first one is that we first rotate the coin register to $\ket{1}$ and then we rotate it back to $\ket{0}$ if the acceptance probability is not 1. This halves the cost, since otherwise one would have to perform 8 multi-controlled rotations (all possible combinations of the control values for registers move, $\phi$ and $\psi$), and in this case we only perform 4 of them. The second simplification again halves the cost, grouping together rotations with similar values. We empirically see that the rotation values controlled on $\ket{000}$ and $\ket{010}$ are very similar, so we group them in rotation $R_0$, that implicitly depends on $\beta$. Similarly, we group the rotations controlled on $\ket{001}$ and $\ket{101}$ in $R_1$, also dependent on $\beta$. This also has the nice effect of only requiring to control on two out of the three bottom registers, ($\phi$, $\psi$ and move), transforming CCC-$R_X$ gates in CC-$R_X$. Separated by the two barriers, from left to right and from top to bottom we can see the implementation of such CC-$R_X$ gates. This figure corresponds to section \ref{sec:Hardware_experiment_results}. $P$ stands for a Phase gate, and $U$ is an $SU(2)$ rotation with the 3 angles indicated.}
\label{fig:coin_flip}\end{figure*}

In section \ref{sec:Hardware_experiment_results} we explain a proof of concept experimental realization of the quantum Metropolis algorithm, and the figure \ref{fig:coin_flip} represents the implementation of the coin flip subroutine of this hardware-adapted algorithm, whose operator is represented by $B$ in \eqref{quantum walk operator tilde W}.

\clearpage
\newpage

\bibliographystyle{ieeetr}
\bibliography{bibliography} 

\end{document}